\documentclass[12pt]{article}
\usepackage[cp1250]{inputenc}
\usepackage[verbose,a4paper,tmargin=0.5in,lmargin=1in,rmargin=1in]{geometry}
\usepackage[pdftex]{graphicx}
\usepackage{amsfonts}
\usepackage{indentfirst}
\usepackage{latexsym}
\usepackage{amsmath}
\usepackage{amssymb}
\usepackage{psfrag}
\usepackage{eufrak}
\usepackage[mathscr]{eucal} 
\usepackage{color}

\newtheorem{Twierdzenie}{Theorem}[section]

\title{On some examples of para-Hermite and para-K\"{a}hler Einstein spaces with $\Lambda \ne 0$.}

\author{$\textrm{Adam Chudecki}^{*}$}

\begin{document}

\maketitle

$*$ Center of Mathematics and Physics, Lodz University of Technology,
\newline
$\ \ \ \ \ $ Al. Politechniki 11, 90-924 Łódź, Poland, adam.chudecki@p.lodz.pl
\newline
\newline
\newline
\textbf{Abstract}. Spaces equipped with two complementary (distinct) congruences of self-dual null strings and at least one congruence of anti-self-dual null strings are considered. It is shown that if such spaces are Einsteinian then the vacuum Einstein field equations can be reduced to a single nonlinear partial differential equation of the second order. Different forms of these equations are analyzed. Finally, several new explicit metrics of the para-Hermite and para-K\"{a}hler Einstein spaces with $\Lambda \ne 0$ are presented. Some relation of that metrics to a modern approach to mechanical issues is discussed.

%#####################################################################################

\section{Introduction}

This paper is devoted to some aspects of complex geometry and real geometry for the neutral signature $(++--)$ in dimension four. 

Complex analysis provides us with a powerful tool in theoretical physics. In relativity complex numbers were used by Einstein himself (he considered imaginary time in special theory of relativity). In the sixties more advanced complex methods, like null tetrad formalism, spinor and twistor formalisms were introduced. It appeared that asymptotically flat spacetime defines 4-dimensional, complex analytic differential manifold equipped with a holomorphic metric. This metric satisfies the vacuum Einstein field equations and the self-dual (SD) or anti-self-dual (ASD) part of its Weyl tensor vanishes. Spaces of such properties are called \textsl{heavenly spaces} (\textsl{$\mathcal{H}$-spaces}). $\mathcal{H}$-spaces also appeared as the nonlinear gravitons in the pioneering work by R. Penrose. They play a fundamental role in twistor theory as well.

In 1976 the idea of \textsl{hyperheavenly spaces} (\textsl{$\mathcal{HH}$-spaces}) was introduced by J.F. Plebański and I. Robinson. $\mathcal{HH}$-spaces are complex spaces which SD (or ASD) part of the Weyl tensor is algebraically degenerate and which satisfy vacuum Einstein field equations with cosmological constant $\Lambda$. In $\mathcal{HH}$-spaces vacuum Einstein  equations have been reduced to a single nonlinear partial differential equation of the second order (\textsl{hyperheavenly equation}). $\mathcal{HH}$-spaces were supposed to be considered as an intermediate step between complex and real theory of relativity. The so called \textsl{Plebański programme} assumed that from the $\mathcal{HH}$-metrics real metrics of Lorentzian signature $(+++-)$ can be obtained. Unfortunately, there were no sufficiently general techniques which allowed one to obtain such solutions. It was the reason why $\mathcal{HH}$-spaces theory became less popular in the nineties.

$\mathcal{HH}$-spaces are equipped with (at least one) family of the totally null and totally geodesic, holomorphic 2-surfaces which are called \textsl{the null strings}. We say that any $\mathcal{HH}$-space admits \textsl{a congruence (foliation) of SD or ASD null strings}. Properties of such congruences have been investigated in \cite{Przanowski, Robinson_Rozga, Plebanski_Rozga}. The relation between the existence of null strings and algebraic structure of the Weyl spinor has been discussed in \cite{Plebanski_Hacyan}. Then the existence of the null strings has been related to the properties of para-Hermite and para-K\"{a}hler spaces. Both these kind of spaces are well-known in physics \cite{Flaherty}.

Real totally null smooth 2-surfaces appeared in \textsl{Walker spaces} as integral manifolds of integrable totally null 2-distributions \cite{Walker}. Walker spaces have been recently investigated in many papers (see \cite{Matsushita, Law_1,Chudecki_1} and references therein). The concept of \textsl{two-sided Walker spaces} was introduced in \cite{Chudecki_1} and then generalized in \cite{Law_1}. In addition, the relation between Walker spaces and $\mathcal{HH}$-spaces was discussed in \cite{Chudecki_1}.

4-dimensional spaces with the metric of neutral signature $(++--)$ play an important role in theoretical physics. They appear in the theory of integrable systems \cite{Dunajski} as spaces which admit ASD conformal structures. Recently, 4-dimensional para-K\"{a}hler Einstein spaces with $\Lambda \ne 0$ have found their place in other geometrical problems, which can be understood as a very modern approach to mechanical issues. In \cite{Bryant_Hsu} the description of the configuration space of two solids rolling on each other without slipping or twisting has been introduced. In \cite{Nurowski_An, Nurowski_Bor_Lamoneda} this approach has been developed. The authors of \cite{Nurowski_An} consider the $(2,3,5)$ distributions over the twistor space $\mathbb{T}(\mathcal{M})$ where $\mathcal{M}$ is a four dimensional manifold equipped with the metric of neutral signature. The surfaces of rolling bodies for which the associated $(2,3,5)$ distribution has the exceptional Lie group $G_{2}$ as a group of local symmetries are considered. Similar problems are analyzed in \cite{Nurowski_Bor_Lamoneda}. The authors pay a special attention to the so called \textsl{dancing metric}, which has many interesting properties. \textsl{Dancing metric} appears to be one of neutral signature and it is equipped with two complementary congruences of SD null 2-surfaces, so it belongs to the para-K\"{a}hler class of metrics. The recent considerations of Nurowski and his collaborators point out even deeper relation between para-K\"{a}hler Einstein spaces with $\Lambda \ne 0$ and $(2,3,5)$ distributions \cite{Nurowski}. 

The unexpected role of such 4-dimensional real manifolds in modern interpretations of mechanical issues motivates us to develop further the results of Ref. \cite{Przanowski_Formanski_Chudecki}. It has been proven in \cite{Przanowski_Formanski_Chudecki} that vacuum Einstein equations in para-Hermite spaces with $\Lambda \ne 0$ can be reduced to a single equation (Eqs. (\ref{Rownanie}) in the present paper). In the case of para-K\"{a}hler Einstein spaces with $\Lambda \ne 0$ we consider the equation (\ref{rownania_Einsteina_Kahler_case}) also presented in \cite{Przanowski_Formanski_Chudecki}. We have been able to find many nontrivial solutions of both these equations. We also use $\mathcal{HH}$-spaces formalism. The corresponding solutions of \textsl{hyperheavenly equation} describe probably the most general solutions of the spaces of the types $[\textrm{D}]^{nn} \otimes [\textrm{II}]^{n}$ and $[\textrm{D}]^{nn} \otimes [\textrm{III,N}]^{e}$ have presented so far (the meaning of the superscripts \textsl{e} and \textsl{n} is explained in subsection \ref{subsubsekcja_congruences_basic}).

The main aim of the present paper is to find explicit examples of para-Hermite and para-K\"{a}hler Einstein spaces with $\Lambda \ne 0$. We do not analyze these metrics in details. The question about the Killing vectors of these metrics is very interesting, but it is outside the scope of this paper. For the same reason, we do not enter the natural question regarding the Lorentzian slices of the metrics of the types $[\textrm{D}]^{ee} \otimes [\textrm{D}]^{ee}$ from the subsection \ref{subsekcja_rozwiazania_DxD_pH} and $[\textrm{D}]^{nn} \otimes [\textrm{D}]^{nn}$ from the subsection \ref{subsekcja_hyperheavenly_DxIID}. 

Our considerations are local and complex in general. We consider complex manifolds of dimension four equipped with the holomorphic metric. The results can be easily carried over to the case of the real manifolds with neutral signature. It is enough to replace all the holomorphic functions and coordinates by the real smooth ones. 

The paper is organized, as follows.

In preliminaries (section \ref{sekcja_Preliminaries}) we remind basic facts about the null tetrad formalism, spinorial formalism, Petrov - Penrose classification and congruences of null strings. In subsection \ref{subsubsection_para-Hermite_spaces} the structure of para-Hermite Einstein spaces is presented. Different approaches to the problem of explicit solutions of the para-Hermite Einstein spaces are sketched. In particular, we explain some of our assumptions taken in the paper.

Section \ref{sekcja_para_Hermite} is devoted to the para-Hermite spaces. Firstly we remind the main result of our previous work \cite{Przanowski_Formanski_Chudecki}: the metric (\ref{metryka}) and the equation (\ref{Rownanie}). Further analysis leads to a large class of exact solutions of this equation. 

In section \ref{para_Kahler_double_null} para-K\"{a}hler Einstein spaces in double null coordinates are considered. The main result of this section is the form of metric generated by (\ref{para_Kahler_nonexpandingASD_nullstrings}). In section \ref{para_Kahler_PRF} we use the Plebański - Robinson - Finley coordinates. Finally, some solutions of different Petrov - Penrose types are given.

Concluding remarks end the paper.

%#####################################################################################

\section{Preliminaries}
\setcounter{equation}{0}
\label{sekcja_Preliminaries}

\subsection{Null tetrad and spinorial formalisms}

In this section we present some basic facts of the null tetrad and spinorial formalisms. As a starting point we consider 4-dimensional complex analytic differentiable manifold endowed with a holomorphic metric $(\mathcal{M},ds^{2})$. The metric $ds^{2}$ can be written in terms of a \textsl{complex null tetrad} $(e^{1}, e^{2}, e^{3}, e^{4})$ as follows
\begin{equation}
ds^{2} = 2  e^{1}  e^{2} + 2 e^{3}  e^{4}
\end{equation}
Let $(\partial_{1}, \partial_{2}, \partial_{3}, \partial_{4})$ be the dual basis of (complex) vectors. The basis $(\partial_{1}, \partial_{2}, \partial_{3}, \partial_{4})$ is also called \textsl{the null tetrad}.

Then we introduce the respective spinorial images of basis of co-vectors and vectors
\begin{equation}
\label{definicccja_gAB}
(g^{A\dot{B}}) := \sqrt{2}
\left[\begin{array}{cc}
e^4 & e^2 \\
e^1 & -e^3
\end{array}\right] 
\ , \ \ \ 
(\partial_{A\dot{B}}) := -\sqrt{2}
\left[\begin{array}{cc}
\partial_{4} & \partial_{2} \\
\partial_{1} & -\partial_{3}
\end{array}\right]  \ , \ \ \ A=1,2 \ , \ \ \ \dot{B}=\dot{1},\dot{2}
\end{equation}
Spinorial indices are manipulated according to the following rules 
\begin{equation}
\label{spinorial_indices_lowering_rule}
m_{A} = \ \in_{A B} m^{B}
\ , \ \ \ 
m^{A} = m_{B} \in^{BA}
\ , \ \ \
m_{\dot{A}} = \ \in_{\dot{A} \dot{B}} m^{\dot{B}}
\ , \ \ \ 
m^{\dot{A}} = m_{\dot{B}} \in^{\dot{B} \dot{A}}
\end{equation}
where $\in_{AB}$ and $\in_{\dot{A}\dot{B}}$ are the spinor Levi-Civita symbols
\begin{eqnarray}
&& (\in_{AB})  := \left[ \begin{array}{cc}
                            0 & 1   \\
                           -1 & 0  
                            \end{array} \right] =:  (\in^{AB} )
\ \ \ , \ \ \ 
 (\in_{\dot{A}\dot{B}})  := \left[ \begin{array}{cc}
                            0 & 1   \\
                           -1 & 0  
                            \end{array} \right] =:  (\in^{\dot{A}\dot{B}} ) 
\\ \nonumber
&& \in_{AC} \in^{AB} = \delta^{B}_{C} \ \ \ , \ \ \ \in_{\dot{A}\dot{C}} \in^{\dot{A}\dot{B}} = \delta^{\dot{B}}_{\dot{C}} \ \ \ , \ \ \ 
(\delta^{A}_{C})= (\delta^{\dot{B}}_{\dot{C}})= \left[ \begin{array}{cc}
                            1 & 0   \\
                            0 & 1  
                            \end{array} \right]
\end{eqnarray}
Define the spin-tensor $g^{aA \dot{B}}$ according to relation $g^{A \dot{B}} = g_{a}^{\ A \dot{B}} \, e^{a}$. Then $g^{aA \dot{B}}$ is a bridge between the null tetrad and spinorial formalisms. It is easy to see that $-\frac{1}{2} g^{aA\dot{B}}g_{bA\dot{B}} = \delta^{a}_{b}$ and $-\frac{1}{2}g^{aA\dot{B}}g_{aC\dot{D}} = \delta^{A}_{C} \delta^{\dot{B}}_{\dot{D}}$. The operators $\partial^{A\dot{B}}$ and $\nabla^{A\dot{B}}$ are the spinorial images of operators $\partial^{a}$ and $\nabla^{a}$, respectively, given by
\begin{equation}
\partial^{A\dot{B}} = g_{a}^{\  A \dot{B}} \partial^{a} \ , \ \ \ 
\nabla^{A\dot{B}} = g_{a}^{\  A \dot{B}} \nabla^{a} 
\end{equation}
In complex relativity the 1-forms $g^{A \dot{B}}$ are unrelated, but real relativity involves some constraints for the 1-forms $g^{A \dot{B}}$. For different signatures of 4-dimensional spaces we find
\begin{eqnarray}
\label{warunki_na_rozna_sygnatury}
(+++-) && \overline{g^{A\dot{B}}} = g^{B \dot{A}}
\\ \nonumber
(++++) && \overline{g^{A\dot{B}}} =- g_{A \dot{B}}
\\ \nonumber
(++--) && \overline{g^{A\dot{B}}} = g^{A \dot{B}}
\end{eqnarray}
where bar stands for the complex conjugation. 

Consider two pairs of normalized spinors $(k_{A}, l_{A})$, $k^{A}l_{A}=1$ and $(k_{\dot{A}}, l_{\dot{A}})$, $k^{\dot{A}} l_{\dot{A}}=1$. They constitute the basis of undotted and dotted 1-index spinors and they generate the new null tetrad as follows
\begin{eqnarray}
\label{transformacje_tetrady}
\widetilde{e}^{\, 1} &:=& \frac{1}{\sqrt{2}} \, k_{A} l_{\dot{B}} \, g^{A \dot{B}}   \\ \nonumber
\widetilde{e}^{\, 2} &:=& \frac{1}{\sqrt{2}} \, l_{A} k_{\dot{B}} \, g^{A \dot{B}}   \\ \nonumber
-\widetilde{e}^{\, 3} &:=& \frac{1}{\sqrt{2}} \, k_{A} k_{\dot{B}} \, g^{A \dot{B}}   \\ \nonumber
\widetilde{e}^{\, 4} &:=& \frac{1}{\sqrt{2}} \, l_{A} l_{\dot{B}} \, g^{A \dot{B}}   
\end{eqnarray}
where $g^{A\dot{B}}$ are given by (\ref{definicccja_gAB}).

For more details about null tetrad and spinorial formalisms, see \cite{Penrose_Rindler, Plebanski_monograph,Plebanski_Przanowski}.

\subsection{Petrov - Penrose classification of the Weyl spinor}

It is well known that every spinor symmetric in all indices can be decomposed according to the formula
\begin{equation} 
\label{decomposition_of_the_arbitrary_spinor}
\Psi_{A_{1}A_{2}...A_{n}} = \Psi_{(A_{1}A_{2}...A_{n})} = \Psi^{(1)}_{(A_{1}} \Psi^{(2)}_{A_{2}}... \Psi^{(n)}_{A_{n})}
\end{equation}
where $\Psi^{(i)}_{A}$ are some 1-index spinors and the bracket $(A_{1}A_{2}...A_{n})$ stands for symmetrization. We use the formula (\ref{decomposition_of_the_arbitrary_spinor}) to classify the spinorial image of the SD part of the Weyl tensor $C_{ABCD} = C_{(ABCD)}$. Firstly, consider the complex case. With complex $C_{ABCD}$ one finds that
\begin{equation}
\label{decomposition_of_SD_Weyl}
C_{ABCD} = \alpha_{(A} \beta_{B} \gamma_{C} \delta_{D)}
\end{equation}
where $\alpha_{A}$, $\beta_{A}$, $\gamma_{A}$ and $\delta_{A}$ are complex spinors. The relations between those spinors brought us to the \textsl{Petrov - Penrose classification of the SD Weyl spinor}
\begin{eqnarray}
\textrm{type [I]} : \ && \ C_{ABCD} = \alpha_{(A} \beta_{B} \gamma_{C} \delta_{D)}  
\\ \nonumber
\textrm{type [II]}: \ && \ C_{ABCD} = \alpha_{(A} \alpha_{B} \beta_{C} \gamma_{D)}   
\\ \nonumber
\textrm{type [D]}: \ && \  C_{ABCD} = \alpha_{(A} \alpha_{B} \beta_{C} \beta_{D)}  
\\ \nonumber
\textrm{type [III]}: \ && \  C_{ABCD} = \alpha_{(A} \alpha_{B} \alpha_{C} \beta_{D)}  
\\ \nonumber
\textrm{type [N]}: \ && \  C_{ABCD} = \alpha_{A} \alpha_{B} \alpha_{C} \alpha_{D} 
\\ \nonumber
\textrm{type }[-]: \ && \  C_{ABCD} = 0 
\end{eqnarray} 
In 4-dimensional spaces with the metric of neutral signature $C_{ABCD}$ is real. It implies a little more detailed Petrov - Penrose classification:
\begin{eqnarray}
\textrm{type [I]}_{r} : \ && \ C_{ABCD} = \alpha_{(A} \beta_{B} \gamma_{C} \delta_{D)}  \ , \ \ \ \alpha_{A}, \beta_{B}, \gamma_{C}, \delta_{D} \textrm{ are real}
\\ \nonumber
 \textrm{type [I]}_{rc} : \ && \ C_{ABCD} = \alpha_{(A} \beta_{B} \gamma_{C} \bar{\gamma}_{D)}  \ , \ \ \ \alpha_{A}, \beta_{B} \textrm{ are real, } \gamma_{C} \textrm{ is complex}
 \\ \nonumber
\textrm{type [I]}_{c} : \ && \ C_{ABCD} = \alpha_{(A} \bar{\alpha}_{B} \beta_{C} \bar{\beta}_{D)}  \ , \ \ \ \alpha_{A}, \beta_{B} \textrm{ are complex} 
 \\ \nonumber
\textrm{type [II]}_{r} : \ && \ C_{ABCD} = \alpha_{(A} \alpha_{B} \beta_{C} \gamma_{D)} \ , \ \ \ \alpha_{A}, \beta_{B}, \gamma_{C} \textrm{ are real}  
\\ \nonumber
\textrm{type [II]}_{rc} : \ && \ C_{ABCD} = \alpha_{(A} \alpha_{B} \beta_{C} \bar{\beta}_{D)} \ , \ \ \ \alpha_{A} \textrm{ is real}, \beta_{B} \textrm{ is complex}  
\\ \nonumber
\textrm{type [D]}_{r} : \ && \  C_{ABCD} = \alpha_{(A} \alpha_{B} \beta_{C} \beta_{D)}  \ , \ \ \ \alpha_{A}, \beta_{B} \textrm{ are real}
\\ \nonumber
\textrm{type [D]}_{c} : \ && \  C_{ABCD} = \alpha_{(A} \alpha_{B} \bar{\alpha}_{C} \bar{\alpha}_{D)}  \ , \ \ \ \alpha_{A} \textrm{ is complex}
\\ \nonumber
\textrm{type [III]} : \ && \  C_{ABCD} = \alpha_{(A} \alpha_{B} \alpha_{C} \beta_{D)}  \ , \ \ \ \alpha_{A}, \beta_{B} \textrm{ are real}
\\ \nonumber
\textrm{type [N]} : \ && \  C_{ABCD} = \alpha_{A} \alpha_{B} \alpha_{C} \alpha_{D} \ , \ \ \ \alpha_{A} \textrm{ is real}
\\ \nonumber
\textrm{type }[-]: \ && \  C_{ABCD} = 0 
\end{eqnarray} 
In \cite{Rod_Hill_Nurowski} the authors used the following symbols for these types: $G_{r}$, $SG$, $G$, $II_{r}$, $II$, $D_{r}$, $D$, $III_{r}$, $N_{r}$ and $0$, respectively. 

We use an abbreviation [deg] which means that Weyl spinor is algebraically degenerated, i.e. it is of the Petrov - Penrose types [II], [D], [III], [N] or $ [-]$ in complex case or of the Petrov - Penrose types $\textrm{[II]}_{r}$, $\textrm{[II]}_{rc}$, $\textrm{[D]}_{r}$, $\textrm{[D]}_{c}$, [III], [N] or $[-]$ in real case. We also use the abbreviation [any] if Weyl spinor is of the arbitrary Petrov - Penrose type.

Five SD curvature coefficients $C^{(i)}$, $i=1,2,3,4,5$ are defined as follows
\begin{eqnarray}
&&C^{(5)} := \ \ \, 2 \, C_{ABCD} \, k^{A}k^{B}k^{C}k^{D} 
\\ \nonumber
&&C^{(4)} := -2 \, C_{ABCD} \, k^{A}k^{B}k^{C}l^{D} 
\\ \nonumber
&&C^{(3)} := \ \ \, 2 \, C_{ABCD} \, k^{A}k^{B}l^{C}l^{D} 
\\ \nonumber
&&C^{(2)} := -2 \, C_{ABCD} \, k^{A}l^{B}l^{C}l^{D} 
\\ \nonumber
&&C^{(1)} := \ \ \, 2 \, C_{ABCD} \, l^{A}l^{B}l^{C}l^{D} 
\end{eqnarray}
where $(k_{A}, l_{A})$, $k^{A}l_{A}=1$ constitute the basis of 1-index undotted spinors. One can always choose the  spinorial basis in such manner, that $k_{A} \sim  \alpha_{A} $. In such \textsl{adapted} coframe (compare (\ref{transformacje_tetrady})) the curvature coefficients read
\begin{eqnarray}
\textrm{type [I]:} \ &&   C^{(5)}=0 \ , \ C^{(4)} \ne 0 \ , \ C^{(3)} \ne 0 \ , \ C^{(2)} \ne 0 \ , C^{(1)} \ne 0
\\ \nonumber
\textrm{type [II]:} \ &&  C^{(5)}=C^{(4)}=0 \ , \ C^{(3)} \ne 0 \ , \ C^{(2)} \ne 0 \ , C^{(1)} \ne 0
\\ \nonumber
\textrm{type [D]:} \ && C^{(5)}=C^{(4)}= 0 \ , \ C^{(3)} \ne 0 \ , \ C^{(2)} \ne 0 \ , C^{(1)} \ne 0 \ 
\\ \nonumber
                     && 2 C^{(2)}C^{(2)}- 3C^{(1)}C^{(3)}=0
\\ \nonumber
\textrm{type [III]:} \ &&  C^{(5)}=C^{(4)}=C^{(3)}=0 \ , \ C^{(2)} \ne 0 \ , C^{(1)} \ne 0
\\ \nonumber
\textrm{type [N]:} \ &&  C^{(5)}=C^{(4)}=C^{(3)}=C^{(2)}=0 \ , \ C^{(1)} \ne 0
\end{eqnarray} 
If additionally $l_{A} \sim \beta_{A}$ then even stronger restrictions on the curvature coefficients appear. They are
\begin{eqnarray}
\textrm{type [I]:} \ &&   C^{(5)}=C^{(1)}=0 \ , \ C^{(4)} \ne 0 \ , \ C^{(3)} \ne 0 \ , \ C^{(2)} \ne 0 
\\ \nonumber
\textrm{type [II]:} \ &&  C^{(5)}=C^{(4)}=C^{(1)}=0 \ , \ C^{(3)} \ne 0 \ , \ C^{(2)} \ne 0 
\\ \nonumber
\textrm{type [D]:} \ && C^{(5)}=C^{(4)}=C^{(2)}=C^{(1)}=0 \ , \ C^{(3)} \ne 0  
\\ \nonumber
\textrm{type [III]:} \ &&  C^{(5)}=C^{(4)}=C^{(3)}= C^{(1)}=0 \ , \ C^{(2)} \ne 0 
\\ \nonumber
\textrm{type [N]:} \ &&  C^{(5)}=C^{(4)}=C^{(3)}=C^{(2)}=0 \ , \ C^{(1)} \ne 0
\end{eqnarray} 
The similar classification holds for the ASD Weyl spinor.

\renewcommand{\arraystretch}{1.2}

\subsection{Congruences of null strings}
\label{subsekcja_Congruences}

\subsubsection{Congruences of null strings - basic concepts}
\label{subsubsekcja_congruences_basic}

In this subsection we collect some basic facts about null strings and their congruences (also called \textsl{foliations}). We equip the manifold $(\mathcal{M}, ds^2)$ with 2-dimensional, holomorphic distribution $\mathcal{D}_{\mu^{A}} = \{ \mu_{A}\nu_{\dot{B}}, \mu_{A} n_{\dot{B}} \}$, $\nu_{\dot{B}}n^{\dot{B}} \ne 0$, defined by the Pfaff system
\begin{equation}
\mu_{A} g^{A \dot{B}}=0
\end{equation}
The distribution $\mathcal{D}_{\mu^{A}}$ is integrable in the Frobenius sense if and only if the spinor $\mu_{A}$ satisfies the equations
\begin{equation}
\label{spinnorowe_rownanie_samodualnej_wstegi}
\mu^{A} \mu^{B} \nabla_{A \dot{B}} \mu_{B} = 0
\end{equation}
Equations (\ref{spinnorowe_rownanie_samodualnej_wstegi}) are called \textsl{SD null strings equations} and they are fundamental in both real and complex geometries equipped with the integrable 2-dimensional distributions \cite{Chudecki_1, Law_1}. If space admits the spinor field $\mu_{A}$ which satisfies the equations (\ref{spinnorowe_rownanie_samodualnej_wstegi}) we say that spinor \textsl{$\mu_{A}$ generates the congruence of SD null strings}. The integral manifolds of the distribution $\mathcal{D}_{\mu^{A}}$ are totally null and geodesic, SD holomorphic 2-surfaces (\textsl{SD null strings}) and they constitute the \textsl{congruence of the SD null strings}. From (\ref{spinnorowe_rownanie_samodualnej_wstegi}) it follows that
\begin{equation}
\label{rozwiniete_rownania_strun_SD}
\nabla_{A \dot{B}} \mu_{B} = Z_{A \dot{B}} \, \mu_{B} + \in_{AB} \! M_{\dot{B}}
\end{equation}
$Z_{A \dot{B}}$ is called the \textsl{Sommers vector} \cite{Sommers} and it is related to the SD part of the conformal curvature. Dotted spinor field $M_{\dot{B}}$ is the \textsl{expansion} of the congruence of SD null strings. With fixed Riemannian structure $(\mathcal{M}, ds^2)$, expansion is the most important property of such congruence and it has a deep geometrical meaning. Indeed, from (\ref{rozwiniete_rownania_strun_SD}) it follows that 
\begin{eqnarray}
M_{\dot{A}}=0  &\Longleftrightarrow &  \nabla_{X}V \in \mathcal{D}_{\mu^{A}} \textrm{ for every vector field } V \in \mathcal{D}_{\mu^{A}} 
\\ \nonumber
&& \textrm{ and for arbitrary vector field } X
\end{eqnarray}
To see this it is enough to contract (\ref{rozwiniete_rownania_strun_SD}) with arbitrary vector field $X^{A\dot{B}}$. One obtains $-2 \nabla_{X} \mu_{B} = X^{A\dot{B}}Z_{A\dot{B}} \, \mu_{B} + X_{B\dot{B}} M^{\dot{B}}$ where $X=-\frac{1}{2} X^{A\dot{B}} \partial_{A\dot{B}}$. Obviously $\nabla_{X} \mu_{B} \sim \mu_{B}$ if and only if $M_{\dot{B}}=0$. Distributions $\mathcal{D}_{\mu^{A}}$ for which $M_{\dot{B}}=0$ are \textsl{parallely propagated}.
 
According to Plebański - Robinson terminology, congruences of the null strings which are parallely propagated $(M_{\dot{B}}=0)$ are called \textsl{nonexpanding} while those for which $M_{\dot{B}} \ne 0$ are called \textsl{expanding}.

[Similarly we can consider congruences of ASD null strings, but they are generated by dotted spinors. ASD null strings generated by the spinor $\nu_{\dot{A}}$ appear as integral manifolds of the distribution $\dot{\mathcal{D}}_{\nu^{\dot{A}}} := \{ \mu_{A} \nu_{\dot{B}} , l_{A} \nu_{\dot{B}} \}$, $\mu_{A} l^{A} \ne 0$.]

The properties of the congruences of null strings in Einstein spaces have been considered in many papers (particularly valuable analysis has been presented in outstanding work of Plebański and Rózga \cite{Plebanski_Rozga}). Here we recall only the most important facts. Especially important is the following theorem
\begin{Twierdzenie}[Generalized Goldberg-Sachs Theorem]
\label{Uogolnione_twierdzenie_Sachsa_Goldberga}
If $(\mathcal{M}, ds^{2})$ is a complex 4-dimensional Einstein space then the following statements are equivalent
\begin{itemize}
\item $(\mathcal{M}, ds^{2})$ admits a congruence of SD null strings generated by the spinor $\mu^{A}$
\item SD Weyl spinor of $(\mathcal{M}, ds^{2})$ is algebraically degenerate and spinor $\mu^{A}$ is a multiple Penrose spinor
\hfill $\blacksquare$ 
\end{itemize}
\end{Twierdzenie}

It follows from Theorem \ref{Uogolnione_twierdzenie_Sachsa_Goldberga} that in Einstein spaces an existence of a congruence of SD (ASD) null strings is equivalent to the algebraic degeneration of the SD (ASD) Weyl spinor. Because of this fact we propose to add to the symbol of algebraically degenerated Petrov - Penrose types additional information about the type of the corresponding congruence. From now on we use the following terminology:
\begin{eqnarray}
\label{terminology_ofthenullstrings}
&& \textrm{superscript } e \textrm{ means that the corresponding congruence is expanding}
\\ \nonumber
&& \textrm{superscript } n \textrm{ means that the corresponding congruence is nonexpanding}
\end{eqnarray}
For example, the symbol $[\textrm{II}]^{e} \otimes [\textrm{III}]^{n}$ means, that the SD Weyl spinor is of the type $[\textrm{II}]$ and ASD Weyl spinor is of the type $[\textrm{III}]$. Moreover, the space is equipped with one expanding congruence of SD null strings and one nonexpanding congruence of ASD null strings. If the superscript is missing it means that the space does not admit any congruence of the SD (ASD) null strings (SD (ASD) type [I]) or such congruence could be expanding or nonexpanding.

Two SD congruences are \textsl{complementary} (\textsl{distinct}, \textsl{transversal}), if and only if spinors $\mu_{A}$ and $\nu_{A}$ which generate those congruences are linearly independent $\mu_{A}\nu^{A} \ne 0$. Analogously, two ASD congruences are complementary, if and only if spinors $\mu_{\dot{A}}$ and $\nu_{\dot{A}}$ which generate those congruences are linearly independent $\mu_{\dot{A}}\nu^{\dot{A}} \ne 0$.

The existence of nonexpanding SD null strings has deep influence on the algebraic type of the SD Weyl spinor. Indeed, we have
\begin{Twierdzenie}
\label{twierdzenie_o_wstegach_nieekspandujacych}
If a complex 4-dimensional Einstein space admits a nonexpanding congruence of SD null strings, then the spinor $\mu_{A}$ which generates this congruence is a multiple Penrose spinor. Moreover
\begin{itemize}
\item $\Lambda \ne 0 \ \Longleftrightarrow \ $ SD Weyl spinor is of the type  [II,D]
\item $\Lambda = 0 \ \Longleftrightarrow \ $ SD Weyl spinor is of the type [III,N,$-$]
\hfill $\blacksquare$ 
\end{itemize}
\end{Twierdzenie}

\subsubsection{Para-Hermite spaces}
\label{subsubsection_para-Hermite_spaces}

Especially interesting are spaces equipped with two complementary congruences of SD (or ASD) null strings. Such spaces are called \textsl{para-Hermite spaces} (see \cite{Flaherty} for more detailed treatment). In what follows we always fix orientation so that the para-Hermite spaces are equipped with two congruences of SD null strings. There is a deep relation between the existence of such congruences and algebraic structure of both SD Weyl spinor and traceless Ricci tensor. This relation deserves a separate paper and will be investigated elsewhere. Here we focus on Einsteinian case. Detailed discussion about the reduction of Einstein equations in Einstein para-Hermite spaces into a single equation has been done in \cite{Przanowski_Formanski_Chudecki}. 

By Theorem \ref{Uogolnione_twierdzenie_Sachsa_Goldberga} it is clear, that the existence of two complementary congruences of SD null strings in Einstein space implies that the SD Weyl spinor can be of the type [D] or [$-$] (left conformally flat space). Congruences can be both expanding or both nonexpanding or of the mixed type (see Table \ref{Tabela_Petrov_Penrose_types}).
\begin{table}[h]
\begin{tabular}{|c|c|c|c|}   \hline
Cosmological & Both congruences & One congruence expanding, &  Both congruences  \\ 
constant & expanding & second nonexpanding  & nonexpanding \\ \hline
$\Lambda \ne 0$ & $\textrm{[D},-]$ & not allowed &  $\textrm{[D]}$ \\ \hline
$\Lambda =0 $ & $\textrm{[D},-]$ & $[-]$  &  $[-]$ \\ \hline
\end{tabular}
\caption{Petrov-Penrose types of the SD Weyl spinor in the para-Hermite Einstein spaces.}
\label{Tabela_Petrov_Penrose_types}
\end{table}

Note that the existence of two complementary congruences of SD null strings of the mixed types in Einstein spaces implies that the SD Weyl spinor vanishes and the space automatically degenerates into left conformally flat space \cite{Przanowski_Formanski_Chudecki}. We are interested in non-conformally flat spaces so in what follows we assume that both SD congruences are expanding or both nonexpanding. According to terminology (\ref{terminology_ofthenullstrings}) we use the following symbols to distinguish SD types [D] equipped with two complementary expanding or two complementary nonexpanding congruences of SD null strings:
\begin{itemize}
\item $[\textrm{D}]^{ee}$ stands for the case when both congruences of SD null strings are expanding
\item $[\textrm{D}]^{nn}$ means that both congruences of SD null strings are nonexpanding
\end{itemize}

If both congruences of SD null strings are nonexpanding, then such spaces become \textsl{para-K\"{a}hler} spaces. For more information about para-K\"{a}hler spaces, see \cite{Flaherty}.

\subsubsection{Approach to the solution problem}
\label{subsubsection_Approach}

To find some nontrivial metrics of the para-Hermite and para-K\"{a}hler Einstein spaces with nonzero cosmological constant $\Lambda$ we use the geometrical structure of complementary congruences of SD null strings. We always choose the null tetrad in such a manner that the distributions which generate those congruences have the following forms
\begin{itemize}
\item $\mathcal{D}_{\mu^{A}} = \textrm{span} (\partial_{1},\partial_{3})$ (the Pfaff system is $e^{2}=0$, $e^{4}=0$)
\item $\mathcal{D}_{\nu^{A}} = \textrm{span} (\partial_{2},\partial_{4})$ (the Pfaff system is $e^{1}=0$, $e^{3}=0$)
\end{itemize}
Then
\begin{eqnarray}
\label{kanoniczny_wybor_tetrady_zerowej}
&& \Gamma_{422}=\Gamma_{424}=0 \ \ \Longleftrightarrow \ \ C^{(5)}=C^{(4)}=0
\\ \nonumber
&& \Gamma_{311}=\Gamma_{313}=0 \ \ \Longleftrightarrow \ \ C^{(2)}=C^{(1)}=0
\end{eqnarray}
The Ricci rotation coefficients $\Gamma_{421}$ and $\Gamma_{423}$ determine the expansion of the congruence defined by the distribution $\mathcal{D}_{\mu^{A}}$, while $\Gamma_{312}$ and $\Gamma_{314}$ determine the expansion of the congruence defined by the distribution $\mathcal{D}_{\nu^{A}}$.

In what follows the null tetrad is always chosen in such a way that (\ref{kanoniczny_wybor_tetrady_zerowej}) is satisfied. Further considerations will prove that the choice of appropriate coordinates is not so natural and different coordinate frames have their own advantages and disadvantages. We use three different coordinate frames. In section \ref{sekcja_para_Hermite} we use the coordinates which have been introduced in \cite{Przanowski_Formanski_Chudecki}. We call them as \textsl{Przanowski coordinates}. In the section \ref{para_Kahler_double_null} so called \textsl{double null coordinates} are used. Finally, \textsl{Plebański - Robinson - Finley coordinates} are introduced in the section \ref{para_Kahler_PRF}.

Then for such null tetrad the vacuum Einstein equations can be always reduced to a single, nonlinear, partial differential equation of the second order. It is the equation (\ref{Rownanie}) for the types $[\textrm{D}]^{ee} \otimes [\textrm{any}]$ with $\Lambda \ne 0$ in Przanowski coordinates or the equation (\ref{rownania_Einsteina_Kahler_case}) for the types $[\textrm{D}]^{nn} \otimes [\textrm{any}]$ with $\Lambda \ne 0$ in double null coordinates. If Plebański - Robinson - Finley coordinates are used, the vacuum Einstein field equations can be reduced to \textsl{expanding} or \textsl{nonexpanding hyperheavenly equation} (for $[\textrm{D}]^{ee} \otimes [\textrm{any}]$ and $[\textrm{D}]^{nn} \otimes [\textrm{any}]$, respectively). 

All these equations are strongly nonlinear and it is very hard to obtain sufficiently general solutions. We are going to solve these equations under additional assumption that the Einstein space admits an ASD null string. Moreover, we will assume that ASD null string has a special form $\textrm{span}(\partial_{1}, \partial_{4})$.

More precisely, we assume the existence of congruence of ASD null strings what is equivalent to the fact that the ASD Weyl spinor is algebraically degenerate (compare Theorem \ref{Uogolnione_twierdzenie_Sachsa_Goldberga}). Algebraic degeneration of the the ASD Weyl spinor implies
\begin{eqnarray}
\label{condition_forthe_degeneration_ofthe_ASDWeyl}
&&    3 \, \dot{C}^{(5)} \dot{C}^{(1)} [\dot{C}^{(3)}]^{4} 
- 2 \, (\dot{C}^{(5)} [\dot{C}^{(2)}]^{2} + \dot{C}^{(1)} [\dot{C}^{(4)}]^{2}) \, [\dot{C}^{(3)}]^{3}  
\\ \nonumber
&& + \frac{2}{3} \, (2 [\dot{C}^{(2)} \dot{C}^{(4)}]^{2} -10 \, \dot{C}^{(5)} \dot{C}^{(4)} \dot{C}^{(2)} \dot{C}^{(1)}
- [\dot{C}^{(5)} \dot{C}^{(1)}]^{2}) \, [\dot{C}^{(3)}]^{2}
\\ \nonumber
&& +2 \, (\dot{C}^{(5)} [\dot{C}^{(2)}]^{2} + \dot{C}^{(1)} [\dot{C}^{(4)}]^{2}) (\dot{C}^{(5)} \dot{C}^{(1)} +2 \, \dot{C}^{(4)} \dot{C}^{(2)} ) \, \dot{C}^{(3)}
\\ \nonumber
&&+ \frac{1}{3} \Big( \dot{C}^{(5)} \dot{C}^{(1)} -4 \, \dot{C}^{(4)} \dot{C}^{(2)} \Big)^{3} 
- \Big( \dot{C}^{(5)} [\dot{C}^{(2)}]^{2} + \dot{C}^{(1)} [\dot{C}^{(4)}]^{2} \Big)^{2}  = 0
\end{eqnarray}
We have two possible ways of further investigations
\begin{enumerate}
\item We can use the tetrad gauge freedom and make $\dot{\mathcal{D}}_{\mu^{\dot{A}}} = \textrm{span} (\partial_{1},\partial_{4})$ what is equivalent to 
\begin{equation}
\label{simmplyfying_assumption}
\dot{C}^{(5)}=\dot{C}^{(4)}=0 \ \Longleftrightarrow \ \Gamma_{411}=\Gamma_{414}=0
\end{equation}
Conditions (\ref{simmplyfying_assumption}) automatically solve (\ref{condition_forthe_degeneration_ofthe_ASDWeyl}). In this case we have to reduce the Einstein vacuum equations from the very beginning, because with such a choice the crucial equations (\ref{Rownanie}), (\ref{rownania_Einsteina_Kahler_case}) or the hyperheavenly equations are, in general, not valid anymore.
\item We can keep the form of the reduced vacuum Einstein equations ((\ref{Rownanie}), (\ref{rownania_Einsteina_Kahler_case}) or hyperheavenly equations) and solve the algebraic condition (\ref{condition_forthe_degeneration_ofthe_ASDWeyl}).
\end{enumerate}

The first way involves some new coordinate frame since the double null coordinates, Przanowski coordinates and Plebański - Robinson - Finley coordinates fail. The second way leaves us with an extremely nasty differential equation (\ref{condition_forthe_degeneration_ofthe_ASDWeyl}) which is so advanced that we have not been able to obtain any solution of it. 

In this point we decided to make an additional \textsl{ad hoc} assumption. We keep the form of the reduced vacuum Einstein field equations and assume that ASD null string is very special: it lies in the plane given by $(\partial_{1},\partial_{4})$. This assumption, which seems to be very strong at the first glance, allows us to find many nontrivial explicit examples of the complex and real metrics of the types $[\textrm{D}]^{ee} \otimes [\textrm{deg}]^{e}$, $[\textrm{D}]^{ee} \otimes [\textrm{deg}]^{n}$, $[\textrm{D}]^{nn} \otimes [\textrm{deg}]^{e}$ and $[\textrm{D}]^{nn} \otimes [\textrm{deg}]^{n}$.

%#####################################################################################

\section{Para-Hermite spaces with two expanding congruences of SD null strings}
\label{sekcja_para_Hermite}
\setcounter{equation}{0}

\subsection{The metric of types $[\textrm{D}]^{ee} \otimes [\textrm{any}]$ with $\Lambda \ne 0$}

The detailed analysis of the spaces of the types $[\textrm{D}]^{ee} \otimes [\textrm{any}]$ equipped with two expanding congruences of SD null strings has been done in \cite{Przanowski_Formanski_Chudecki} where the procedure of reduction of vacuum Einstein field equations led the authors to a suitable coordinate system $(x,y,z,t)$ (\textsl{Przanowski coordinates}). Then it has been proven that the metric of the spaces of types $[\textrm{D}]^{ee} \otimes [\textrm{any}]$ can be brought to the following form
\begin{eqnarray}
\label{metryka}
ds^{2}& = & \dfrac{1}{2v_{x}}\left(v_{x}dx - dt + v_{y}dy - v_{z}dz\right)
\left(v_{x}dx + dt - v_{y}dy + v_{z}dz\right)  
\\ \nonumber
 & &  +2x^{2}v_{x}\, \mbox{\large $e$}^{\frac{Q_{x}}{\kappa_{x}}}
dydz
\end{eqnarray}
where
\begin{equation}
\label{defffinija_funckji_v}
v(x,y,z):=Q-\kappa \, \frac{Q_{x}}{\kappa_{x}} \ , \ \ \ \kappa(x):=\left(\frac{\gamma_0}{x^2}+\frac{\Lambda x}{3}\right)^{-1}
\end{equation}
$\gamma_{0}$ is some constant and $\Lambda$ is cosmological constant. (We use the notation $v_{x} := \frac{\partial v}{\partial x}$, $v_{y} := \frac{\partial v}{\partial y}$, etc.). The metric (\ref{metryka}) admits the Killing vector field $\partial_{t}$. The function $Q=Q(x, y, z)$ has to satisfy the crucial equation
\begin{equation}
\label{Rownanie}
Q_{yz}+x^{2}\kappa_{x}%\,
(\mbox{\large $e$}^{\frac{Q_{x}}{\kappa_{x}}})_{x}
+2\kappa\, \mbox{\large $e$}^{\frac{Q_{x}}{\kappa_{x}}}=0.
\end{equation}

Now we choose the null tetrad in the form
\begin{eqnarray}
\label{wybor_tetrady}
&& e^{1} = \frac{1}{2} \, d(v+t)
\\ \nonumber
&& e^{2} = \frac{1}{2 v_{x}} \, d(v-t) - \frac{v_{z}}{v_{x}} \, dz
\\ \nonumber
&& e^{3} = - dy
\\ \nonumber
&& e^{4} = \frac{v_{y}}{2 v_{x}} \, d(v-t) - \bigg( \frac{v_{y}v_{z}}{v_{x}} + F \bigg) \, dz
\end{eqnarray}
where 
\begin{equation}
F := x^{2}  v_{x} \exp \Big( \frac{Q_{x}}{\kappa_{x}} \Big)
\end{equation}
From the first structure equations the connection forms can be extracted as
\begin{eqnarray}
&& \Gamma_{42} = \frac{1}{x} \, e^{3} \ , \ \ \ \Gamma_{31} = - \frac{v_{y}}{xv_{x}} \, e^{2} + \frac{1}{xv_{x}} \, e^4
\\ \nonumber
&& \Gamma_{41} = \Sigma \, e^{1} + \Omega \, e^{3} \ , \ \ \ \Gamma_{32} = \bigg( v_{y} \Omega - \frac{\partial}{\partial x} \Big( \frac{v_{y}}{v_{x}} \Big) \bigg) e^{1} + \frac{1}{x} \, e^{4} + B \, e^{3}
\\ \nonumber
&& \Gamma_{12} + \Gamma_{34} = -\frac{1}{x} \, e^{2} + \bigg( \frac{1}{\kappa} - \frac{1}{xv_{x}} \bigg) e^{1} + \bigg( \frac{F_{y}}{F} - \frac{v_{y}}{xv_{x}} - \frac{v_{yx}}{v_{x}} + \frac{v_{y}}{\kappa} \bigg) e^{3}
\\ \nonumber
&& -\Gamma_{12} + \Gamma_{34} = \bigg( \frac{1}{xv_{x}} - \frac{1}{\kappa} - 2\Omega  \bigg) e^{1} - \frac{1}{x} \, e^{2} - A \, e^{3}
\end{eqnarray}
where
\begin{subequations}
\begin{eqnarray}
\label{definicje_pewnych_wielkosci}
&& \Sigma := \frac{v_{zx}}{F v_{x}^{2}} 
\\
\label{definicja_Omegi}
&& \Omega := \frac{1}{xv_{x}} - \frac{1}{F} \frac{\partial}{\partial z} \Big( \frac{v_{y}}{v_{x}} \Big)
\\ 
&& A := \frac{v_{y}}{xv_{x}} - \frac{\partial}{\partial x} \Big( \frac{v_{y}}{v_{x}} \Big) - \frac{2 v_{y}}{F} \frac{\partial}{\partial z} \Big( \frac{v_{y}}{v_{x}} \Big) + \frac{F_{x}v_{y}}{F v_{x}} - \frac{F_{y}}{F}
\\ \nonumber
&& \ \ \ \ \ \equiv
2 v_{y} \Omega - 2 \frac{\partial}{\partial x} \Big( \frac{v_{y}}{v_{x}} \Big) +  
\frac{v_{y}}{xv_{x}}  - \frac{Q_{y}}{\kappa}
\\ 
&& B := v_{x} \frac{\partial}{\partial y} \Big( \frac{v_{y}}{v_{x}} \Big) - v_{y} \frac{\partial}{\partial x} \Big( \frac{v_{y}}{v_{x}} \Big) - \frac{v_{y}^{2}}{F} \frac{\partial}{\partial z} \Big( \frac{v_{y}}{v_{x}} \Big) + \frac{F_{x} v_{y}^{2}}{F v_{x}} - \frac{F_{y}v_{y}}{F}
\\ \nonumber
&& \ \ \ \ \ \equiv
v_{y} A - v_{y}^{2} \Omega + v_{x} \frac{\partial}{\partial y} \Big( \frac{v_{y}}{v_{x}} \Big)
\end{eqnarray}
\end{subequations}
After quite long calculations we get the conformal curvature in the form
\begin{eqnarray}
&& C^{(1)}=C^{(2)}=C^{(4)}=C^{(5)}=0 \ , \ \ \ C^{(3)} = 2 \gamma_{0} \, x^{-3}
\\ \nonumber
&& \dot{C}^{(5)} = - \frac{2}{F} \Sigma_{z} \ , \ \ \ \dot{C}^{(4)} = \frac{2}{F} \Omega_{z} \equiv 2 \Sigma_{x} + \frac{2v_{y}}{F} \Sigma_{z} + \frac{2}{x} \Sigma
\\ \nonumber
&& \dot{C}^{(3)} = -2 \Omega_{x} - \frac{2v_{y}}{F} \Omega_{z} - \frac{2}{x} \Omega  - \frac{2}{3} \Lambda \equiv - \frac{A_{z}}{F} + \frac{\Omega}{x} + \frac{1}{x^{2} v_{x}} + \frac{\Lambda}{3}
\\ \nonumber
&& \dot{C}^{(2)} = A_{x} + \frac{v_{y}}{F} A_{z} + \frac{v_{y}}{x} \Omega  - \frac{1}{x} \frac{\partial}{\partial x} \Big( \frac{v_{y}}{v_{x}} \Big) \equiv 
\frac{2B_{z}}{F} - \frac{A}{x} - \frac{1}{x} \left( \frac{F_{y}}{F} + \frac{v_{y}}{\kappa} + \frac{v_{y}}{x v_{x}} - \frac{v_{yx}}{v_{x}} \right)
\\ \nonumber
&& \ \ \ \ \ \ \equiv -2 v_{y} \dot{C}^{(3)} - v_{y}^{2} \dot{C}^{(4)} + 2 v_{yx} \Omega - \frac{\Lambda}{3} v_{y} - \frac{\partial}{\partial x} \bigg( \frac{Q_{y}}{\kappa} + 2 \frac{\partial}{\partial x}  \Big(  \frac{v_{y}}{v_{x}}  \Big)     \bigg)
\\ \nonumber
&& \dot{C}^{(1)} =  - 2B_{x} - \frac{2v_{y}}{F} B_{z}  \equiv -3v_{y} \dot{C}^{(2)} -3 v_{y}^{2} \dot{C}^{(3)} - v_{y}^{3} \dot{C}^{(4)} 
\\ \nonumber
 && \ \ \ \ \ \ \ \ -2 \frac{\partial}{\partial x} \bigg( v_{x} \frac{\partial}{\partial y} \Big( \frac{v_{y}}{v_{x}} \Big) \bigg) - 2 v_{yx} \bigg( \frac{v_{y}}{xv_{x}} - \frac{Q_{y}}{\kappa} - 2 \frac{\partial}{\partial x} \Big( \frac{v_{y}}{v_{x}} \Big) \bigg)
\end{eqnarray}
and, of course, the traceless Ricci tensor vanishes $C_{ab}=0$ and the curvature scalar is $R=-4\Lambda$.

\subsection{Coordinate gauge freedom}

The tetrad (\ref{wybor_tetrady}) is chosen in such a manner that planes $(\partial_{1}, \partial_{3})$ and $(\partial_{2}, \partial_{4})$ constitute the congruences of the SD null strings. Coordinate transformations should maintain these planes, so it is reasonable to consider as a starting point the rotations $(\partial_{1}, \partial_{3}) \rightarrow (\partial'_{1}, \partial'_{3})$ and $(\partial_{2}, \partial_{4}) \rightarrow (\partial'_{2}, \partial'_{4})$. Nevertheless, some portion of this freedom has been used earlier during the process of reduction of the metric to the form (\ref{metryka}) and the vacuum Einstein equations to the equation (\ref{Rownanie}). After some work we find that the tetrad (\ref{wybor_tetrady}) remains invariant under the transformations 
\begin{equation}
\label{transf_coordinates}
t' = \frac{1}{c_{0}} t  +f(y) - g(z) \ , \ \ \   x' = c_{0}x \ , \ \ \ 
y' = y'(y) \ , \ \ \ z'=z'(z)
\end{equation}
Under these transformations the constant $\gamma_{0}$ and functions $v$ and $Q$ become
\begin{equation}
\label{transf_functions}
\gamma'_{0} = c_{0}^{3} \gamma_{0} \ , \ \ \ 
v' = \frac{1}{c_{0}} v +f(y) +g(z) \ , \ \ \ Q' = \frac{1}{c_{0}} Q - \frac{\kappa(x)}{c_{0}} \, \ln \left( \frac{ dy'}{dy} \frac{dz'}{dz} \right) +f(y) +g(z)
\end{equation}
In above formulas $f$, $g$, $y'$ and $z'$ are arbitrary functions of their variables and $c_{0}$ is an arbitrary constant. Under (\ref{transf_coordinates}) the null tetrad transforms as follows
\begin{eqnarray}
&& e'^{1} = \frac{1}{c_{0}} \, e^{1} - f_{y} \, e^{3} \ \ \ \ \ \ \  \ \ \ \ \ \ \   \ \ \ \ \ \ \ \partial'_{1} = c_{0} \, \partial_{1}
\\ \nonumber
&& e'^{2} = c_{0} \, e^{2} \ \ \ \ \ \ \   \ \ \ \ \ \ \  \ \ \ \ \ \ \ \ \ \  \ \ \ \ \ \ \  \partial'_{2} = \frac{1}{c_{0}}  \, \partial_{2} - f_{y}  \, \partial_{4}
\\ \nonumber
&& e'^{3} = y'_{y} \, e^{3}  \ \ \ \ \ \ \  \ \ \ \ \ \ \  \ \ \ \ \ \ \ \ \ \ \ \ \ \ \ \ \  \partial'_{3} = \frac{c_{0}f_{y}}{y'_{y}} \, \partial_{1} + \frac{1}{y'_{y}} \, \partial_{3}
\\ \nonumber
&& e'^{4} = \frac{c_{0}f_{y}}{y'_{y}} \, e^{2} + \frac{1}{y'_{y}} \, e^{4}  \ \ \ \ \ \ \   \ \ \ \  \ \ \ \ \ \ \ \partial'_{4} = y'_{y}  \, \partial_{4}
\end{eqnarray}

\subsection{Special class of solutions of the types $[\textrm{D}]^{ee} \otimes  [\textrm{deg}]$}

To find the solutions of Eq. (\ref{Rownanie}) we first assume that (\ref{simmplyfying_assumption}) is satisfied. This implies (see sec. \ref{subsubsection_Approach})
\begin{equation}
\Sigma=0 
\end{equation}
Then, according to (\ref{definicje_pewnych_wielkosci}) one finds that $v_{zx}=0$. The general solution of this condition gives the solution for the $Q$ function 
\begin{equation}
\label{rozwiazanie_na_algebraiczna_degeneracje}
Q(x,y,z) = h(y,z) + \kappa(x) l(y,z) + m(x,y)
\end{equation}
Putting (\ref{rozwiazanie_na_algebraiczna_degeneracje}) into (\ref{Rownanie}) we obtain the system of three equations
\begin{subequations}
\begin{eqnarray}
\label{equation_1}
&&l_{yz} = N(y) \, e^l 
\\
\label{equation_2}
&& h_{yz} = G(y) \, e^{l} 
\\
\label{equation_3}
 &&x^2 \kappa_{x} \, \partial_{x} \big( e^{\frac{m_{x}}{\kappa_{x}}} \big) +2\kappa e^{\frac{m_{x}}{\kappa_{x}}}+ N(y) \kappa +G(y)=0
\end{eqnarray} 
\end{subequations}
with some arbitrary functions $N(y)$ and $G(y)$. Equation (\ref{equation_3}) can be easily solved. One gets
\begin{equation}
e^{\frac{m_{x}}{\kappa_{x}}}=S(y) \, \frac{\Lambda x^3 -6\gamma_0}{x}-G(y) \, \frac{2\Lambda x^3 -3\gamma_0}{6x^2}-\frac{N(y)}{2}
\end{equation}
where $S(y)$ is the integration function. So
\begin{eqnarray}
m(x,y)&=&\int \kappa_x \, \ln R(x,y)  dx = \kappa  \, \ln R(x,y) - \int \kappa \, (\ln R(x,y))_{x}  dx = 
\\ \nonumber
&=&\kappa  \, \ln R(x,y) - \int \frac{6xS(y) -  G(y)}{xR(x,y)} dx
\end{eqnarray}
where 
\begin{equation}
\label{definicja_R}
R(x,y) := S(y) \, \frac{\Lambda x^3 -6\gamma_0}{x}-G(y) \, \frac{2\Lambda x^3 -3\gamma_0}{6x^2}-\frac{N(y) }{2}
\end{equation}
Then the $Q$ function reads 
\begin{equation}
\label{soluttion_for_QQ}
Q(x,y,z) = h(y,z) + \kappa(x) l(y,z) + \kappa(x) \ln R(x,y) - \int \frac{6xS(y) -  G(y)}{xR(x,y)} dx 
\end{equation}
Consequently, the function $v$ defined by (\ref{defffinija_funckji_v}) is
\begin{equation}
\label{soluttion_for_vv}
v(x,y,z) = h(y,z)  - \int \frac{6xS(y) -  G(y)}{xR(x,y)} dx 
\end{equation} 
Since necessarily $v_{x} \ne 0$, $S(y)$ and $G(y)$ cannot vanish at the same time
\begin{equation}
\label{warunek_na_poprawnosc_metryki}
|S(y)| + |G(y)| \ne 0
\end{equation}
The functions $h(y,z)$ and $l(y,z)$ in (\ref{soluttion_for_QQ}) and (\ref{soluttion_for_vv}) have to satisfy the equations (\ref{equation_1}) and (\ref{equation_2}). To solve these equations we need some additional assumption about $N(y)$. [Indeed, cases $N=0$ and $N \ne 0$ have to be considered separately].
\newline
\textbf{Important Note:} If we demand that the congruence of ASD null strings be nonexpanding, then we have to impose the condition $\Gamma_{413}=\Omega=0$. Using (\ref{soluttion_for_QQ}) and (\ref{soluttion_for_vv}) in (\ref{definicja_Omegi}) we quickly arrive at the formula
\begin{equation}
\Omega = - \frac{6xS R}{ (6xS-G)^{2} }
\end{equation}
Obviously $\Omega=0 \ \Longleftrightarrow \ S(y)=0$. The case with nonexpanding congruence of ASD null string can be identified by putting $S(y)=0$.

Straightforward, but somewhat long calculations bring us to the nonzero ASD curvature coefficients 
\begin{eqnarray}
\label{krzywizna_dla_ansatza}
&&\dot{C}^{(3)} = \frac{2 (\Lambda G^{3} + 18 GNS + 648 \gamma_{0} S^{3})}{3 (6xS - G)^{3}}
\\ \nonumber
&&\dot{C}^{(2)} = -3 v_{y} \dot{C}^{(3)}  -x v_{x}^{2} \frac{\partial}{\partial y} 
\bigg( \frac{R^{2} (6xS+G)}{(6xS - G)^{3}}    \bigg)
\\ \nonumber
&& \ \ \ \ \ \ = -3 v_{y} \dot{C}^{(3)} + \frac{3}{xR(6xS-G)^{2}} \Big[ \big( 12S (S N_{y} - S_{y} N) + 2\Lambda G (S G_{y} - S_{y} G) \big) x^{2} 
\\ \nonumber
&& \ \ \ \ \ \ \ \ \ +  4N (SG_{y}  - S_{y} G)  x 
 + 36\gamma_{0} S (S G_{y} - S_{y} G) + \frac{1}{3} G (NG_{y} - G N_{y}) \Big]
 \\ \nonumber
&&\dot{C}^{(1)} = -4 v_{y} \dot{C}^{(2)} - 6 v_{y}^{2} \dot{C}^{(3)} - 2F v_{x}^{2} \frac{\partial}{\partial y} \bigg(  \frac{v_{xy}}{F v_{x}^{2}}  \bigg)
\\ \nonumber
&& \ \ \ \ \ \ = -4 v_{y} \dot{C}^{(2)} - 6 v_{y}^{2} \dot{C}^{(3)} + \frac{2e^{l} (6xS-G)^{3}}{xR^{2}} \frac{\partial}{\partial y} \bigg(  \frac{T}{e^{l}(6xS-G)^{3}}  \bigg)
\end{eqnarray}
where we put
\begin{equation}
\label{definicja_T}
-xR^{2} v_{xy} = \frac{1}{x}(SG_{y} - GS_{y})(\Lambda x^{3} + 3 \gamma_{0}) + 3x(SN_{y} - NS_{y}) + \frac{1}{2} (NG_{y} - GN_{y}) =: T
\end{equation}

To find explicit solutions we have to split the problem into two cases. Since, generally, we are not interested in left conformally flat spaces we assume $\gamma_{0} \ne 0$. Firstly, consider the case when $N(y) \ne  0$. If $N(y) \ne 0$ then Eq. (\ref{equation_1}) is the Liouville equation which the general solution is well known as
\begin{equation}
\label{sollution_for_l}
l(y,z)=\ln \left(\frac{2 r_{y}\tilde{r}_{z}}{N(y)(r+\tilde{r})^2} \right)
\end{equation}
where $r(y)$ and $\tilde{r}(z)$ are arbitrary functions of their variables. Putting (\ref{sollution_for_l}) into (\ref{equation_2}) we obtain
\begin{equation}
\label{sollution_for_h}
h(y,z) = -2 \int \frac{G(y)}{N(y)} \frac{\partial}{\partial y} \ln(r+\tilde{r}) dy + w(y) + \tilde{w} (z)
\end{equation}
Using the coordinate gauge freedom (\ref{transf_coordinates}) and (\ref{transf_functions}) one can always put $r(y)=y$, $\tilde{r}(z)=z$, $w(y)=\tilde{w}(z)=0$ without any loss of generality . Finally, the solution is
\begin{equation}
\label{OGOLNE_solution_1}
Q(x,y,z) = -2 \int \frac{G(y)}{N(y)} \frac{dy}{y+z} + \kappa(x)  \ln \left(\frac{2 }{N(y)(y+z)^2} \right) + \int \kappa_{x} \, \ln R(x,y) dx
\end{equation}
Assume now the case $N(y) =  0$. If $N(y)=0$ then $l_{yz}=0$, so $l(y,z) = \ln(r_{y} \tilde{r}_{z})$, where $r=r(y)$ and $\tilde{r} = \tilde{r}(z)$. Using the coordinate gauge freedom we can put $r(y)=y$, $\tilde{r}(z)=z$ without any loss of generality. Consequently, $l(y,z)=0$. Solution for $h(y,z)$ reads
\begin{equation}
h(y,z) = z \int G(y) dy + w(y) + \tilde{w}(z)
\end{equation}
Functions $w$ and $\tilde{w}$ can be gauged away, as before. Consequently the general solution for the case when $N(y)=0$ takes the form of
\begin{equation}
\label{OGOLNE_solution_2}
Q(x,y,z) = z \int G(y) dy  + \int \kappa_{x} \, \ln R(x,y) dx
\end{equation}

Gathering, we arrive at the following metric of the type $[\textrm{D}]^{ee}  \otimes [\textrm{deg}]$
\begin{equation}
\label{metryka_ogolna_alenienajlepsza}
ds^{2} = - \frac{6xS-G}{2xR} dx^{2} + \frac{xR}{2(6xS-G)} \theta^{2} - 2x(6xS-G)e^{l} dy dz
\end{equation}
where 
\begin{eqnarray}
 \theta &:=& dt + h_{z} dz - \Big( h_{y} - \frac{\partial}{\partial y} \int \frac{(6xS-G)dx}{xR} \Big) dy
\\ \nonumber
&\equiv& dt + h_{z} dz - \Big( h_{y} - \int \frac{T dx}{xR^{2}} \Big) dy
\end{eqnarray}
$R$ and $T$ are given by (\ref{definicja_R}) and (\ref{definicja_T}), respectively; $S$, $G$ and $N$ are arbitrary functions of $y$ only. Moreover
\begin{eqnarray}
\nonumber
\textrm{if $N=0$:} && e^{l}=1 \ , \ \ \ h=z \int Gdy
\\ \nonumber
\textrm{if $N \ne 0$:} && e^{l}= \frac{2}{N(y+z)^{2}}\ , \ \ \ h=-2 \int \frac{G}{N} \frac{dy}{y+z}
\end{eqnarray}
The curvature is given by (\ref{krzywizna_dla_ansatza}). 

There are three arbitrary functions $N$, $S$ and $G$ involved in our considerations. However, it appears that the number of arbitrary functions in the metric (\ref{metryka_ogolna_alenienajlepsza}) can be reduced to two functions only: $S$ and $G$ (if $N=0$) or $S/N$ and $G/N$ (if $N\ne0$). In the next subsection we take advantage of this fact and we are going to find slightly more suitable forms of the metric.

\subsection{Metrics of the types $[\textrm{D}]^{ee} \otimes [\textrm{II}]^{e}$ and $[\textrm{D}]^{ee} \otimes [\textrm{II}]^{n} $}
\label{subsekcja_rozwiazania_DxII_pH}

\subsubsection{Type $[\textrm{D}]^{ee} \otimes [\textrm{II}]^{e}$ and $[\textrm{D}]^{ee} \otimes [\textrm{II}]^{n} $ (Class 1)}
\label{subsubsekcja_typ_II_class_1}

ASD type [II] is characterized by $\dot{C}^{(3)} \ne 0$ and $\tau \ne 0$ (compare (\ref{warunek_na_typ_D})). After some work we arrive at the two classes of solutions of the type $[\textrm{D}]^{ee} \otimes [\textrm{II}]$. First class of these solutions is characterized by $N \ne 0$. Define $f (y) := S/N$ and $g(y) := G/N$. With these abbreviations we obtain the metric
\begin{equation}
\label{metrykka_typu_DxII_1}
ds^{2} = - \frac{6xf-g}{2xr} dx^{2} + \frac{xr}{2(6xf-g)} \theta^{2} - \frac{4x(6xf-g)}{(y+z)^2} dy dz
\end{equation}
where
\begin{equation}
\theta := dt +h_{z} dz - ( h_{y} - F ) dy
\end{equation}
and
\begin{eqnarray}
r(x,y) &:=& f \Big( \Lambda x^2 - \frac{6 \gamma_{0}}{x} \Big) - g \Big( \frac{\Lambda x}{3} - \frac{\gamma_{0}}{2x^{2}} \Big) - \frac{1}{2}
\\ \nonumber
h(y,z) &:=& -2 \int \frac{g(y) dy}{y+z}
\\ \nonumber
F(x,y) &:=& \frac{\partial}{\partial y} \int \frac{(6xf-g)dx}{xr}
\end{eqnarray}
The curvature reads
\begin{eqnarray}
&&\dot{C}^{(3)} = \frac{2(\Lambda g^3 +18gf+ 648\gamma_{0} f^3)}{3 (6xf-g)^3}
\\ \nonumber
&&\dot{C}^{(2)} = -3 (h_{y}-F) \dot{C}^{(3)} + \frac{3}{xr(6xf-g)^2} \bigg( \big( 2 \Lambda g (g_{y}f-gf_{y}) -12ff_{y} \big) x^2 + 
\\ \nonumber
&& \ \ \ \ \ \ \ \ \ \ \ \ \ \ \ \ \ \ \ \ \ \ \ \  \ \ \ \ \ \ \ \ \ \ \  +4(fg_{y}-gf_{y}) x + 36\gamma_{0} f(fg_{y} - gf_{y}) + \frac{1}{3}g g_{y}    \bigg)
\\ \nonumber
&&\dot{C}^{(1)} = -4(h_{y}-F) \dot{C}^{(2)} - 6(h_{y}-F)^2 \dot{C}^{(3)}
\\ \nonumber
&& \ \ \ \ \ \ \ \ \ \ \ \ \ \ \  + \frac{4 (6xf-g)^3}{(y+z)^{2}xr^2}
 \frac{\partial}{\partial y} \bigg(\frac{ (y+z)^{2} \big(  (fg_{y}-gf_{y}) (\Lambda x^2 + \frac{3 \gamma_{0}}{x}  ) -3xf_{y} + \frac{1}{2}g_{y} \big) }{ 2(6xf-g)^{3}}  \bigg)
\end{eqnarray}
where $f$ and $g$ are arbitrary functions of $y$ such that $|f_{y}| + |g_{y}| \ne 0$, otherwise the solution degenerates into the type $[\textrm{D}]^{ee} \otimes [\textrm{D}]$. Moreover, there must be $\Lambda g^3 +18gf+ 648\gamma_{0} f^3 \ne 0$, since otherwise the solution degenerates into the type  $[\textrm{D}]^{ee} \otimes [\textrm{III}]$. 

Note that nonvanishing or vanishing of the function $f$ distinguishes the cases with expanding or nonexpanding, respectively, congruence of ASD null strings. Indeed, we have
\begin{eqnarray}
\nonumber
&& f \ne 0 \ \Longleftrightarrow \ \textrm{the type } [\textrm{D}]^{ee} \otimes [\textrm{II}]^{e}
\\ \nonumber
&& f = 0 \ \Longleftrightarrow \ \textrm{the type } [\textrm{D}]^{ee} \otimes [\textrm{II}]^{n}
\end{eqnarray}

\subsubsection{Type $[\textrm{D}]^{ee} \otimes [\textrm{II}]^{e}$ (Class 2)}

This is a class of solutions with $N=0$. Here we have $S\ne0$, so the congruence of ASD null strings is necessarily expanding. Changing abbreviations for the functions and using the gauge freedom we arrive at the metric
\begin{equation}
\label{metrykka_typu_DxII_2}
ds^{2} = - \frac{6xf-1}{2xr} dx^{2} + \frac{xr}{2(6xf-1)} \theta^{2} - 2x(6xf-1) dy dz
\end{equation}
with
\begin{equation}
 \theta := dt + y dz - ( z - F ) dy
\end{equation}
where 
\begin{eqnarray}
r(x,y) &:=& f \Big( \Lambda x^2 - \frac{6 \gamma_{0}}{x} \Big) - \frac{\Lambda x}{3} + \frac{\gamma_{0}}{2x^{2}}
\\ \nonumber
F(x,y) &:=& \frac{\partial}{\partial y} \int \frac{(6xf-1)dx}{xr}
\end{eqnarray}
Here $f=f(y)$ is an arbitrary function such that $f_{y} \ne 0$, otherwise the solution degenerates into type $[\textrm{D}]^{ee} \otimes [\textrm{D}]^{ee}$. However, if $\Lambda = 0$ then for of the same reason there must be $f \ne (a_{0} y +b_{0})^{-1/2}$. Curvature reads
\begin{eqnarray}
\dot{C}^{(3)} &=& \frac{2(\Lambda + 648\gamma_{0} f^3)}{3 (6xf-1)^3}
\\ \nonumber
\dot{C}^{(2)} &=& -3 (z-F) \dot{C}^{(3)} - \frac{3(2\Lambda f_{y} x^2 + 36\gamma_{0} f f_{y})}{xr(6xf-1)^2}
\\ \nonumber
\dot{C}^{(1)} &=& -4(z-F) \dot{C}^{(2)} - 6(z-F)^2 \dot{C}^{(3)} - \frac{2 (6xf-1)^3 (\Lambda x^3 + 3 \gamma_{0})}{x^2r^2} \frac{\partial}{\partial y} \bigg(\frac{f_{y}}{ (6xf-1)^{3}}  \bigg)
\end{eqnarray}

\subsection{Metrics of the type $[\textrm{D}]^{ee} \otimes [\textrm{D}]^{ee}$ or $[\textrm{D}]^{ee} \otimes [\textrm{D}]^{nn}$}
\label{subsekcja_rozwiazania_DxD_pH}

\subsubsection{$\tau$-condition}

To distinguish the types $[\textrm{D}]^{ee} \otimes [\textrm{II}]$ and $[\textrm{D}]^{ee} \otimes [\textrm{D}]$ with (\ref{simmplyfying_assumption}) fulfilled we need to consider the invariant 
\begin{equation}
\label{warunek_na_typ_D}
\tau := 2 \dot{C}^{(2)} \dot{C}^{(2)} - 3 \dot{C}^{(3)} \dot{C}^{(1)}
\end{equation}
Type $[\textrm{D}]^{ee} \otimes [\textrm{II}]$ degenerates into $[\textrm{D}]^{ee} \otimes [\textrm{II}]$ if $\tau=0$ and, of course, $\dot{C}^{(3)} \ne 0$. The condition $\tau=0$ is equivalent to the equation
\begin{eqnarray}
\label{warunek_na_typ_DD}
&&9\Big[ \big( 12S (S N_{y} - S_{y} N) + 2\Lambda G (S G_{y} - S_{y} G) \big) x^{2} 
\\ \nonumber
&& \ \ \ \ \ \ +  4N (SG_{y}  - S_{y} G)  x 
 + 36\gamma_{0} S (S G_{y} - S_{y} G) + \frac{1}{3} G (NG_{y} - G N_{y}) \Big]^{2}
\\ \nonumber
&& -2x (\Lambda G^{3} + 18 NGS + 648\gamma_{0} S^{3}) \Big( (T_{y}- T l_{y}) (6xS-G) - 3T (6xS_{y} - G_{y}) \Big) = 0
\end{eqnarray}
where $T$ is defined by (\ref{definicja_T}). The left hand side of the condition (\ref{warunek_na_typ_DD}) is a polynomial in $x$ and it seems to be rather involved. But after thorough analysis we arrive at the two possibilities:
\begin{eqnarray}
(i) \ \ && N\ne 0 \ , \ \ \ G=G_{0} N \ , \ \ \ S=S_{0} N \ , \ \ \ 
\\ \nonumber
&& |S_{0}| + |G_{0}| \ne 0 \ , \ \ \ \Lambda G_{0}^{3} + 18G_{0} S_{0} + 648 \gamma_{0} S_{0}^{3} \ne 0
\end{eqnarray}
where $S_{0}$ and $G_{0}$ are arbitrary constants.
\begin{eqnarray}
\label{Type_DxD_solution_class_2b}
(ii) \ \ && N=0 \ , \ \ \ S G_{y} - S_{y} G = S^{3} e^{l} b(z) \ , \ \ \ \Lambda S b=0
\\ \nonumber
&& |S(y)| + |G(y)| \ne 0 \ , \ \ \ \Lambda G^{3} + 648 \gamma_{0} S^{3} \ne 0
\end{eqnarray}
Where $b=b(z)$ is an arbitrary function of its coordinate. The function $l$ can be gauged to zero, which makes $b = b_{0} = \textrm{const}$.

\subsubsection{Types $[\textrm{D}]^{ee} \otimes [\textrm{D}]^{ee}$ and $[\textrm{D}]^{ee} \otimes [\textrm{D}]^{nn}$ (Class 1)}

Here the solution can be obtained from the solution described in \ref{subsubsekcja_typ_II_class_1} by putting $f=S_{0} = \textrm{const}$ and $g=G_{0} = \textrm{const}$. Namely, we obtain the metric
\begin{equation}
\label{Metryka_DxD_clasa1}
ds^{2} = - \frac{6xS_{0}-G_{0}}{2xr} dx^{2} + \frac{xr}{2(6xS_{0}-G_{0})} \theta^{2} - \frac{4x(6xS_{0}-G_{0})}{(y+z)^2} dy dz
\end{equation}
where
\begin{equation}
 \theta := dt - \frac{2G_{0}}{y+z} dz + \frac{2G_{0}}{y+z} dy
\end{equation}
and
\begin{equation}
r(x) := S_{0} \Big( \Lambda x^2 - \frac{6 \gamma_{0}}{x} \Big) - G_{0} \Big( \frac{\Lambda x}{3} - \frac{\gamma_{0}}{2x^{2}} \Big) - \frac{1}{2}
\end{equation}
The curvature reads
\begin{eqnarray}
&&\dot{C}^{(3)} = \frac{2(\Lambda G_{0}^3 +18S_{0}G_{0}+ 648\gamma_{0} S_{0}^3)}{3 (6xS_{0}-G_{0})^3}
\\ \nonumber
&&\dot{C}^{(2)} =  \frac{6G_{0}}{y+z} \dot{C}^{(3)}
\\ \nonumber
&&\dot{C}^{(1)} = \frac{24G_{0}^{2}}{(y+z)^{2}} \dot{C}^{(3)}
\end{eqnarray}
The solution depends on four constants $S_{0}$, $G_{0}$, $\gamma_{0}$ and $\Lambda$, such that $|S_{0}|+|G_{0}| \ne 0$. 
To keep the type $[\textrm{D}]^{ee} \otimes [\textrm{D}]$ there must be $\Lambda G_{0}^3 +18S_{0}G_{0}+ 648\gamma_{0} S_{0}^3 \ne 0$, otherwise the solution degenerates into the type $[\textrm{D}]^{ee} \otimes [\textrm{III}]$. Note, that
\begin{eqnarray}
\nonumber
&& S_{0} \ne 0 \ \Longleftrightarrow \ \textrm{the type } [\textrm{D}]^{ee} \otimes [\textrm{D}]^{ee}
\\ \nonumber
&& S_{0} = 0 \ \Longleftrightarrow \ \textrm{the type } [\textrm{D}]^{ee} \otimes [\textrm{D}]^{nn}
\end{eqnarray}
If $S_{0}=0$ the metric (\ref{Metryka_DxD_clasa1}) is of the type $[\textrm{D}]^{ee} \otimes [\textrm{D}]^{nn}$ and it is equipped with four congruences of null strings. SD congruences are expanding, while ASD congruences are both nonexpanding (compare Theorem \ref{twierdzenie_o_wstegach_nieekspandujacych} and the Table \ref{Tabela_Petrov_Penrose_types}). Note that such metrics do not admit any Lorentzian slices, so they cannot be treated as a special cases of Plebański - Demiański solution \cite{Plebanski_Demianski}. To the best of our knowledge this is the first example of the vacuum type $[\textrm{D}] \otimes [\textrm{D}]$ complex space-time with $\Lambda \ne 0$ which does not admit any Lorentzian real section.

\subsubsection{Types $[\textrm{D}]^{ee} \otimes [\textrm{D}]^{ee}$ and $[\textrm{D}]^{ee} \otimes [\textrm{D}]^{nn}$ (Class 2a)}

The detailed analysis of (\ref{Type_DxD_solution_class_2b}) proves that this class divides into two subclasses. If $S_{y}G-SG_{y} =0$ we arrive at the metric
 \begin{equation}
\label{metryka_typu_DD_class2_a}
ds^{2} = - \frac{6xS_{0}-G_{0}}{2xr} dx^{2} + \frac{xr}{2(6xS_{0}-G_{0})} \theta^{2} - 2x(6xS_{0}-G_{0}) dy dz
\end{equation}
where
\begin{equation}
 \theta := dt + G_{0}y \, dz -G_{0} z \,  dy
\end{equation}
and
\begin{equation}
r(x) := S_{0} \Big( \Lambda x^2 - \frac{6 \gamma_{0}}{x} \Big) - G_{0} \Big( \frac{\Lambda x}{3} - \frac{\gamma_{0}}{2x^{2}} \Big) 
\end{equation}
The curvature is given by
\begin{eqnarray}
&&\dot{C}^{(3)} = \frac{2(\Lambda G_{0}^3 + 648\gamma_{0} S_{0}^3)}{3 (6xS_{0}-G_{0})^3}
\\ \nonumber
&&\dot{C}^{(2)} = -3zG_{0} \, \dot{C}^{(3)}
\\ \nonumber
&&\dot{C}^{(1)} = 6z^{2} G_{0}^{2}\, \dot{C}^{(3)}
\end{eqnarray}
The solution depends on four constants $S_{0}$, $G_{0}$, $\gamma_{0}$ and $\Lambda$ and constants $G_{0}$ and $S_{0}$ must fulfill  $|S_{0}|+|G_{0}| \ne 0$ and $\Lambda G_{0}^3 + 648\gamma_{0} S_{0}^3 \ne 0$. Significance of the $S_{0}$ constant is related, as before, to the properties of the ASD null strings. We find
\begin{eqnarray}
\nonumber
&& S_{0} \ne 0 \ \Longleftrightarrow \ \textrm{the type } [\textrm{D}]^{ee} \otimes [\textrm{D}]^{ee}
\\ \nonumber
&& S_{0} = 0 \ \Longleftrightarrow \ \textrm{the type } [\textrm{D}]^{ee} \otimes [\textrm{D}]^{nn}
\end{eqnarray}
It follows that the case with nonexpanding congruences of ASD null strings corresponds to the $S_{0}=0$. For $S_{0}=0$ the metric (\ref{metryka_typu_DD_class2_a}) does not admit Lorentzian slices. 

[\textbf{Remark:} note, that para-Hermite spaces of the type $[\textrm{D}]^{ee} \otimes [\textrm{D}]^{nn}$ are also para-K\"{a}hler spaces.]

\subsubsection{Type $[\textrm{D}]^{ee} \otimes [\textrm{D}]^{ee}$ (Class 2b)}

The second subclass is characterized by $S_{y}G - S G_{y} \ne 0$. It involves $\Lambda =0$, so the solution is not so interesting for our purposes. Nevertheless, we present it here for the completeness. We have $S \ne 0$ and $S_{y}G - S G_{y} = b_{0} S^{3}$. After some work we get the metric
 \begin{equation}
ds^{2} = - \frac{(6x-a_{0}-b_{0}y)x}{\gamma_{0} (a_{0}+b_{0}y-12x)} dx^{2} + \frac{\gamma_{0} (a_{0}+b_{0}y-12x)}{4x (6x-a_{0}-b_{0}y)} \theta^{2} - 2x(6x-a_{0}-b_{0}y) dy dz
\end{equation}
where
\begin{eqnarray}
\theta & := &dt + \Big( a_{0}y + \frac{1}{2}b_{0}y^{2} \Big) \, dz - z (a_{0} + b_{0}y) \,  dy
\\ \nonumber
&& - \frac{b_{0}}{144 \gamma_{0}} \bigg( \frac{24x(6x-a_{0}-b_{0}y)}{a_{0}+b_{0}y-12x} -2(a_{0}+b_{0}y) \ln (a_{0}+b_{0}y -12x)  \bigg)  dy
\end{eqnarray}
The curvature reads
\begin{eqnarray}
&&\dot{C}^{(3)} = \frac{432 \gamma_{0}}{(6x-a_{0}-b_{0}y)^3}
\\ \nonumber
&&\dot{C}^{(2)} = 3 (a_{0}+b_{0}y) \bigg( \frac{b_{0}}{72 \gamma_{0}} \ln(a_{0}+b_{0}y-12x) -z   \bigg) \, \dot{C}^{(3)}
\\ \nonumber
&&\dot{C}^{(1)} = 6 (a_{0}+b_{0}y)^{2} \bigg( \frac{b_{0}}{72 \gamma_{0}} \ln(a_{0}+b_{0}y-12x) -z   \bigg)^{2} \, \dot{C}^{(3)}
\end{eqnarray}
The solution depends on three constants $a_{0}$, $b_{0}$ and $\gamma_{0}$. Congruences of ASD null strings are here necessarily expanding.

\subsection{Metrics of the types $[\textrm{D}]^{ee} \otimes [\textrm{III}]^{e}$ and $[\textrm{D}]^{ee} \otimes [\textrm{III}]^{n}$}
\label{subsekcja_rozwiazania_DxIII_pH}

\subsubsection{Type $[\textrm{D}]^{ee} \otimes [\textrm{III}]$ conditions}

The type $[\textrm{D}]^{ee} \otimes [\textrm{III}]$ is determined by the conditions $\dot{C}^{(3)}=0$ and $\dot{C}^{(2)} \ne 0$. The condition $\dot{C}^{(3)}=0$ implies that
\begin{equation}
\label{condition_for_type_III}
\Lambda G^{3} + 18 GNS + 648 \gamma_{0} S^{3} = 0
\end{equation}
Some remarks should now be given:
\begin{itemize}
\item $G \ne 0$, otherwise from (\ref{condition_for_type_III}) and from $\gamma_{0} \ne 0$ it follows that $S=0$. However, $G$ and $S$ cannot vanish at the same time.
\item $N \ne 0$, otherwise from (\ref{condition_for_type_III}) it follows that $S$ is proportional to $G$ or $S=0$, which implies $\dot{C}^{(2)}=0$
\item Functions $S$, $G$ and $N$ cannot be simultaneously proportional to each other because it causes $\dot{C}^{(2)}=0$.
\end{itemize} 
Now we are ready to solve the condition (\ref{condition_for_type_III}) explicitly. We find two families of solutions
\newline
\textbf{Class 1} (expanding congruence of the ASD null strings)
\begin{equation}
S \ne 0 \ , \ \ \ N= - \frac{36 \gamma_{0} S^{2}}{G} - \frac{\Lambda G^{2}}{18S} \ , \ \ \ S_{y}G - SG_{y} \ne 0
\end{equation}
\textbf{Class 2} (nonexpanding congruence of the ASD null strings)
\begin{equation}
S = 0 \ , \ \ \ \Lambda=0 \ , \ \ \ N_{y}G - NG_{y} \ne 0
\end{equation}

\subsubsection{Type $[\textrm{D}]^{ee} \otimes [\textrm{III}]^{e}$}

Here we have $S \ne 0$, which corresponds to the existence of expanding congruence of the ASD null strings. We denote $f(y) := S /G$ where $f$ is an arbitrary function of $y$ such that $f_{y} \ne 0$. The metric reads
\begin{equation}
\label{metryka_DxIII_class1}
ds^{2} = - \frac{6xf-1}{2xfr} dx^{2} + \frac{xfr}{2(6xf-1)} \theta^{2} + \frac{ 4x(6xf-1)}{(y+z)^{2} \big( 36\gamma_{0} f^2 + \frac{\Lambda}{18f} \big)} dy dz
\end{equation}
where
\begin{eqnarray}
\theta &:=& dt + h_{z} dz - (h_{y}  - F )dy
\end{eqnarray}
and 
\begin{eqnarray}
r(x,y) &:=& (6xf-1)^{2} \bigg( \frac{\Lambda}{36f^2} + \frac{\gamma_{0}}{2x^2 f} \bigg)
\\ \nonumber
h(y,z) &:=& 2 \int \frac{dy}{(y+z) \Big( 36\gamma_{0} f^2 + \frac{\Lambda}{18f} \Big) }
\\ \nonumber
F(x,y) &:=& \frac{\partial}{\partial y} \bigg[ \frac{1}{36\gamma_{0}f^2 + \frac{\Lambda }{18f}} \bigg( \ln \frac{(6xf-1)^{2}}{1+ \frac{\Lambda x^{2}}{18f \gamma_{0}}} 
+ \frac{12f}{\sqrt{\frac{\Lambda}{18f\gamma_{0}}}} \textrm{atan} \Big( \sqrt{\frac{\Lambda}{18f 
\gamma_{0}}}  x \Big)
  \bigg) \bigg]
\end{eqnarray}
Curvature reads
\begin{eqnarray}
&&\dot{C}^{(2)} = - \Big( 36\gamma_{0} f + \frac{\Lambda}{18f^2} \Big) \frac{f_{y}}{xrf}
\\ \nonumber
&&\dot{C}^{(1)} = -4 (h_{y} -F) \dot{C}^{(2)} 
\\ \nonumber
&&
\ \ \ \ \ \ \ \ \ \ 
+ \frac{2 (6xf-1)^{3}}{ (y+z)^{2} xr^{2}f^{2} \big( 36\gamma_{0}f^{2} + \frac{\Lambda}{18f} \big) } \frac{\partial}{\partial y}
\bigg(  \frac{(y+z)^{2} xrf^{2} \big( \frac{\Lambda}{36f^{2}} + \frac{3 \gamma_{0}}{x} \big) \dot{C}^{(2)}}{(6xf-1)}  \bigg)
\end{eqnarray}

\subsubsection{Type $[\textrm{D}]^{ee} \otimes [\textrm{III}]^{n}$}

This solution corresponds to $S=0=\Lambda$, so the congruence of the SD null strings is nonexpanding here. We introduce the function $f(y) := G/N$ such that $f_{y} \ne 0$. After some work we obtain the metric
\begin{equation}
\label{metryka_DxIII_class2}
ds^{2} =  \frac{f}{xr} dx^{2} - \frac{xr}{4f} \theta^{2} + \frac{4xf}{(y+z)^2} dy dz
\end{equation}
where
\begin{equation}
 \theta := dt +h_{z} dz + \Big( -h_{y} +  \frac{f_{y}(r+1)}{r} + f_{y} \ln (x^{2}r) \Big) dy
\end{equation}
and
\begin{eqnarray}
r(x,y) &:=& \frac{f \gamma_{0}}{x^{2}} - 1
\\ \nonumber
h(y,z) &:=& -2 \int \frac{f(y) dy}{y+z}
\end{eqnarray}
The curvature is
\begin{eqnarray}
&&\dot{C}^{(2)} = \frac{2f_{y}}{xfr}
\\ \nonumber
&&\dot{C}^{(1)} = 2 \dot{C}^{(2)}  \bigg( \frac{2f(2r+1)}{(y+z)r}  + \frac{f f_{yy}}{r f_{y}} + 2f_{y} \ln (x^{2}r) + \frac{f_{y}(2r-1)}{r}     \bigg)
\end{eqnarray}

\subsection{Solutions of the types $[\textrm{D}]^{ee} \otimes [\textrm{N},-]$}

For the types $[\textrm{D}]^{ee} \otimes [\textrm{N},-]$ we have $\dot{C}^{(3)}=\dot{C}^{(2)}=0$ and this is equivalent to the following conditions
\begin{eqnarray}
&& \Lambda G^{3} + 18 GNS + 648 \gamma_{0} S^{3} = 0
\\ \nonumber
&& N(S_{y} G - S G_{y}) = 0
\\ \nonumber
&& G (GN_{y} - N G_{y}) + 108 \gamma_{0} S (GS_{y} - S G_{y}) = 0
\\ \nonumber
&& 6S (NS_{y} - S N_{y}) + \Lambda G (G S_{y} - S G_{y}) = 0
\end{eqnarray}
Here, like before, we obtain two classes of solutions
\newline
\textbf{Class 1}
\begin{equation}
 N\ne 0 \ , \ \ \ G=G_{0} N \ , \ \ \ S=S_{0} N \ , \ \ \ 
 \Lambda G_{0}^{3} + 18G_{0} S_{0} + 648 \gamma_{0} S_{0}^{3} =0
\end{equation}
where $S_{0}$ is an arbitrary constant and $G_{0}$ is arbitrary nonzero constant.
\newline
\textbf{Class 2}
\begin{equation}
 N= 0  \ , \ \ \ 
  G \ne 0 \ , \ \ \ S = - \Big( \frac{\Lambda}{648\gamma_{0}} \Big)^{\frac{1}{3}} G
\end{equation}
Then it can be easily found, that for both these classes $T=0$ (compare (\ref{definicja_T})). Conditions $\dot{C}^{(3)}=\dot{C}^{(2)}=0$ imply $\dot{C}^{(1)}=0$. It means that using our solution in the form (\ref{rozwiazanie_na_algebraiczna_degeneracje}) we cannot obtain the examples of the types $[\textrm{D}]^{ee} \otimes [\textrm{N}]$. The type $[\textrm{D}]^{ee} \otimes [\textrm{III}]$ degenerates to the type $[\textrm{D}]^{ee} \otimes[-] $.

%#####################################################################################

\section{Para-K\"{a}hler Einstein spaces in double null coordinates}
\label{para_Kahler_double_null}
\setcounter{equation}{0}

\subsection{The metric of para-K\"{a}hler spaces}

In this section we consider the Einstein para-Hermite spaces which are equipped with two nonexpanding congruences of SD null strings, i.e., the spaces of the types $[\textrm{D}]^{nn} \otimes [\textrm{any}]$. Such spaces are called \textsl{para-K\"{a}hler}. It has been proven in \cite{Przanowski_Formanski_Chudecki} that the metric of such spaces can be brought to the following form 
\begin{equation}
\label{para_hermite_dwiewstegi_nonexpanding}
ds^2 = 2 (M_{px} \, dp + M_{qx} \, dq) dx +2 (M_{py} \, dp   + M_{qy} \, dq) dy
\end{equation}
where $(p,q,x,y)$ are local coordinates and the function $M=M(p,q,x,y)$ is such that
\begin{equation}
f:= M_{px}M_{qy} - M_{qx}M_{py} \ne 0
\end{equation}
We introduce the null tetrad
\begin{eqnarray}
\label{para_Kahler_null_tetrad}
&& e^{1}=dx \ , \ \ \ e^{2} = M_{px} \, dp + M_{qx} \, dq
\\ \nonumber
&& e^{3}=dy \ , \ \ \ e^{4} = M_{py} \, dp + M_{qy} \, dq
\end{eqnarray}
and the dual basis is
\begin{eqnarray}
&& \partial_{1} = \partial_{x} \ , \ \ \ \partial_{2} = \frac{M_{qy}}{f} \, \partial_{p} - \frac{M_{py}}{f} \, \partial_{q}
\\ \nonumber
&& \partial_{3} = \partial_{y} \ , \ \ \ \partial_{4} = -\frac{M_{qx}}{f} \, \partial_{p} + \frac{M_{px}}{f} \, \partial_{q}
\end{eqnarray}
The connection forms read now
\begin{eqnarray}
&& \Gamma_{42}=\Gamma_{31}=0 \ , \ \ \ \Gamma_{41} = A e^{1} + B e^{3} \ , \ \ \ \Gamma_{32} = C e^{1} + D e^{3}
\\ \nonumber
&& \Gamma_{12}+\Gamma_{34} = -(\ln f)_{x} \, e^{1} - (\ln f)_{y} \, e^{3} 
\\ \nonumber
&&-\Gamma_{12}+\Gamma_{34} = \big( (\ln f)_{x}-2B \big) e^{1} - \big( (\ln f)_{y}+2C \big) e^{3} 
\end{eqnarray}
in which we used the abbreviations
\begin{eqnarray}
&& fA := M_{px}M_{qxx} - M_{qx}M_{pxx} \ , \ \ \ fB := M_{px}M_{qxy} - M_{qx}M_{pxy}
\\ \nonumber
&& fC := M_{py}M_{qxy} - M_{qy}M_{pxy} \ , \ \ \ fD := M_{py}M_{qyy} - M_{qy}M_{pyy}
\end{eqnarray}
Congruences of the SD null strings are spanned by the vectors $(\partial_{1}, \partial_{3})$ and $(\partial_{2}, \partial_{4})$. Of course, both these congruences are nonexpanding: $\Gamma_{42}=\Gamma_{31}=0$.

From the second Cartan structure equations we find the conformal curvature as
\begin{eqnarray}
&& C^{(5)}=C^{(4)}=C^{(2)}=C^{(1)}=0
\\ \nonumber
&& 3fC^{(3)} = \frac{f}{2}R = M_{qy} (\ln f)_{xp} -M_{py} (\ln f)_{xq} -M_{qx} (\ln f)_{yp} +M_{px} (\ln f)_{qy}
\\ \nonumber
&& \dot{C}^{(5)} = 2 \partial_{4} A = \frac{2}{f} (\partial_{q} (M_{px}A) - \partial_{p}(M_{qx}A))
\\ \nonumber
&& \dot{C}^{(4)} =  \partial_{2} A - \partial_{4} B = \frac{1}{f} \partial_{p}( M_{qy}A + M_{qx}B ) 
 - \frac{1}{f} \partial_{q} ( M_{py}A + M_{px}B )
\\ \nonumber
&&\dot{C}^{(3)} = \frac{1}{3} \partial_{2} (\partial_{x} f - 3B)+ \frac{1}{3} \partial_{4} (\partial_{y} f +3C)
= \frac{1}{3f} \partial_{p} \big( (\partial_{x} \ln f - 3B) M_{qy} - (\partial_{y} \ln f + 3C) M_{qx} \big)
\\ \nonumber
&& \ \ \ \ \ \ \ \ 
+\frac{1}{3f} \partial_{q} \big((\partial_{y} \ln f + 3C) M_{px} - (\partial_{x} \ln f - 3B) M_{py} \big)
\\ \nonumber
&& \dot{C}^{(2)} = \partial_{2}C - \partial_{4}D = \frac{1}{f} \partial_{p}( M_{qy}C + M_{qx}D ) 
 - \frac{1}{f} \partial_{q} ( M_{py}C + M_{px}D )
\\ \nonumber
&& \dot{C}^{(1)} = -2 \partial_{2}D = \frac{2}{f} ( \partial_{q}(M_{py}D) - \partial_{p} (M_{qy}D) )
\end{eqnarray}
and the trace-less Ricci tensor is given by
\begin{eqnarray}
&& C_{11}=C_{31}=C_{33} = C_{44}=C_{42}=C_{22}=0
\\ \nonumber
&&2f C_{12} = -2fC_{34} = M_{qy} (\ln f)_{px}- M_{py} (\ln f)_{qx}+ M_{qx} (\ln f)_{py}- M_{px} (\ln f)_{qy}
\\ \nonumber
&& fC_{32} =  M_{qy} (\ln f)_{py} - M_{py} (\ln f)_{qy}
\\ \nonumber
&& fC_{41} =  M_{px} (\ln f)_{qx} - M_{qx} (\ln f)_{px}
\end{eqnarray}
The metric (\ref{para_hermite_dwiewstegi_nonexpanding}) admits the following coordinate gauge freedom
\begin{eqnarray}
\label{coordinate_gauge_freedom_Kahler_case}
&& x'=x'(x,y) \ , \ \ \ y'=y'(x,y) \ , \ \ \ p'=p'(p,q) \ , \ \ \ q'=q'(p,q) 
\\ \nonumber
&& M'=M +a(x,y) + b(p,q)
\end{eqnarray}
(where $a$ and $b$ are arbitrary functions of their variables).

Now, consider the case of para-K\"{a}hler Einstein spaces. It is quite easy to see that the conditions $C_{ab}=0$ and $R=-4\Lambda$ imply that the function $M$ must satisfy the crucial equation
\begin{equation}
\label{rownania_Einsteina_Kahler_case}
f:= M_{px}M_{qy} - M_{qx}M_{py} = e^{-\Lambda M}
\end{equation}
Some portion of freedom (\ref{coordinate_gauge_freedom_Kahler_case}) has been used to reduce the Einstein field equations to the form (\ref{rownania_Einsteina_Kahler_case}). For Einstein para-K\"{a}hler spaces the functions $a$ and $b$ are not arbitrary anymore. They are defined by
\begin{equation}
e^{\Lambda a} = \frac{\partial x'}{\partial x} \frac{\partial y'}{\partial y} - \frac{\partial x'}{\partial y} \frac{\partial y'}{\partial x}
\ , \ \ \ 
e^{\Lambda b} = \frac{\partial p'}{\partial p} \frac{\partial q'}{\partial q} - \frac{\partial p'}{\partial q} \frac{\partial q'}{\partial p}
\end{equation}

\subsection{Para-K\"{a}hler Einstein spaces with algebraically special ASD Weyl spinor}

\subsubsection{Simplifying assumption}

The condition for algebraic degeneration of the ASD Weyl spinor appears to be a rather nasty equation for the $M$ function. As before we assume that the ASD null strings are somewhat special and they are spanned by the vectors $(\partial_{1}, \partial_{4})$. It is equivalent to the condition (\ref{simmplyfying_assumption}). Straightforward calculations show that assuming (\ref{simmplyfying_assumption}) we are left with the set of equations
\begin{subequations}
\begin{eqnarray}
\label{warunki_przy_istnieniu_ASD_struny}
M_{q} = r \, M_{p} + s  && (\textrm{consequence of the condition } \Gamma_{411}=0)
\\ 
\label{consequences_of_Cdot4zero}
rr_{py}-r_{qy}-2r_{y}r_{p} - r_{y} \Lambda s = 0 && (\textrm{consequence of the condition } \dot{C}^{(4)}=0) \ \ \ \ \ \
\\ 
\label{Einstein_equations_for_Dxany_non}
M_{px}(r_{y}M_{p} + s_{y}) = e^{-\Lambda M} && (\textrm{Einstein equations})
\end{eqnarray}
\end{subequations}
where $r=r(p,q,y)$ and $s=s(p,q,y)$ are arbitrary functions of their variables. The Ricci rotation coefficient $\Gamma_{413} = B$ describes the expansion of the congruence of ASD null strings. Using (\ref{warunki_przy_istnieniu_ASD_struny}) we find
\begin{equation}
B=r_{y} M_{px}^{2} e^{\Lambda M}
\end{equation}
Now it is clear that $r_{y}$ has a deep geometrical interpretation: $r_{y} \ne 0$ means that the congruence of ASD null strings is expanding, while $r_{y}=0$ implies that the expansion of this congruence vanishes. With $\Gamma_{411}=0$ the coordinate transformations (\ref{coordinate_gauge_freedom_Kahler_case}) are restricted to the  transformations with $y'=y'(y)$.

\subsubsection{Nonexpanding case}
\label{subsubsekcja_nonexpanding_case}

Firstly we deal with the nonexpanding case, $r_{y}=0$. Note that with $\Lambda \ne 0$ the only types which admit the nonexpanding  congruences of ASD null strings are ASD types [II] and [D] (see Theorem \ref{twierdzenie_o_wstegach_nieekspandujacych}). Using the gauge freedom for the function $p'=p'(p,q)$ one proves that the function $r(p,q)$ can be brought to zero without any loss of generality. The function $M$ reads then $M(x,y,p,q) = l(y,p,q) + g(x,y,p)$. Putting this form of $M$ into Einstein  equation (\ref{rownania_Einsteina_Kahler_case}) we find that this equation separates into two Liouville equations
\begin{equation}
\label{dwa_rownania_Liouvillea}
g_{xp}e^{\Lambda g} = \frac{1}{h} \ , \ \ \ l_{yq}e^{\Lambda l} = h 
\end{equation}
where $h=h(p,y)$ is an arbitrary function of its variables. Eqs. (\ref{dwa_rownania_Liouvillea}) can be easily solved. Then the general solution for $M$ reads
\begin{equation}
M=\frac{1}{\Lambda} \ln \left( \frac{\Lambda^{2} (\alpha-\beta)^{2} (\gamma-\delta)^{2}}{4\alpha_{q}\beta_{y}\gamma_{x}\delta_{p}} \right) 
\end{equation}
with $\alpha=\alpha(p,q)$, $\beta=\beta(y,p)$, $\gamma=\gamma(x,y)$ and $\delta=\delta(y,p)$ being arbitrary functions. However, gauge freedom still available is strong enough to bring the functions $\alpha$ and $\gamma$ to especially simple form, namely, $\alpha=q$, $\gamma=x$.

Gathering, the large class of para-K\"{a}hler metrics (\ref{para_hermite_dwiewstegi_nonexpanding}) with $\Lambda \ne 0$ which admits a  nonexpanding congruence of ASD null strings and is of the ASD type [II] or [D] is generated by the function $M$ of the form
\begin{equation}
\label{para_Kahler_nonexpandingASD_nullstrings}
M=\frac{1}{\Lambda} \ln \left( \frac{\Lambda^{2} (q-\beta)^{2} (x-\delta)^{2}}{4\beta_{y}\delta_{p}} \right) 
\end{equation}
which depends on two arbitrary functions $\beta=\beta(y,p)$ and $\delta=\delta(y,p)$. Note, that this metric remains unchanged under the coordinate gauge freedom
\begin{eqnarray}
\label{gauge_freedom_po_uproszczeniach}
&& x'=x+\tilde{x}(y) \ , \ \ \ q'=q+\tilde{q}(p) \ , \ \ \ y'=y'(y) \ , \ \ \ p'=p'(p) 
\\ \nonumber
&& \beta'=\beta + \tilde{q}(p) \ , \ \ \ \delta'=\delta + \tilde{x}(y)
\end{eqnarray}
with $\tilde{x}$ and $\tilde{q}$ being arbitrary gauge functions. After rather long and tedious calculations we find the ASD curvature coefficients
\begin{eqnarray}
\label{ASD_curvature_nonexpanding}
  \dot{C}^{(3)} &=& -\frac{2}{3} \Lambda  \ , \ \ \   \dot{C}^{(5)}=\dot{C}^{(4)}=0
\\ \nonumber
 f \dot{C}^{(2)} &=& -\frac{4\beta_{y}}{\Lambda (q-\beta)^{2}} \left( \frac{\partial}{\partial y} \frac{\delta_{pp}}{\delta_{p}} + \frac{2\delta_{yp}}{x-\delta} + \frac{2\delta_{p}\delta_{y}}{(x-\delta)^{2}} \right)
\\ \nonumber
 \frac{f}{2} \dot{C}^{(1)} &=& \frac{(x-\delta)^{2}}{(q-\beta)^{3}} \frac{2\beta_{y}^{2}}{\Lambda \delta_{p}} \frac{\partial}{\partial y} \left( \frac{\partial}{\partial p} \frac{\delta_{pp}}{\delta_{p}} - \frac{1}{2}  \frac{\delta_{pp}^{2}}{\delta_{p}^{2}}   \right)
- \frac{1}{x-\delta} \frac{16\delta_{py}\delta_{y}\beta_{y}}{\Lambda (q-\beta)^{2}}
 - \frac{1}{(x-\delta)^{2}} \frac{8\delta_{p}\delta_{y}^{2}\beta_{y}}{\Lambda (q-\beta)^{2}}
\\ \nonumber
&&- \frac{(x-\delta)^{2}}{(q-\beta)^{2}} \frac{\beta_{y}}{\Lambda} \left(  \frac{1}{ \delta_{p}} \partial_{y}\partial_{p} \ln (\beta_{y}\delta_{p}) \partial_{y}\partial_{p} \ln (\delta_{p}) +
 \frac{\partial}{\partial p} \left( \frac{\beta_{y}}{\delta_{p}} \partial_{y} \Big( \frac{1}{\beta_{y}} \partial_{y}\partial_{p} \ln (\beta_{y}\delta_{p}) \Big)  \right)  \right)
 \\ \nonumber
 && + \frac{x-\delta}{(q-\beta)^{2}} \frac{2\beta_{y}}{\Lambda} \frac{\partial}{\partial p} \left( \frac{\partial}{\partial y}  \frac{\beta_{yy}}{\beta_{y}}  - 2  \frac{\delta_{py}^{2}}{\delta_{p}^{2}}  - \frac{1}{2}  \frac{\beta_{yy}^{2}}{\beta_{y}^{2}}  \right) - \frac{1}{(q-\beta)^{2}} \frac{8\beta_{y}}{\Lambda} \frac{\partial}{\partial p} \left( \frac{\delta_{y}\delta_{py}}{\delta_{p}} \right)
\end{eqnarray}
Solutions given by (\ref{para_Kahler_nonexpandingASD_nullstrings}) are in general of the ASD type [II] which degenerates to the type [D] if $\tau=0$ (see (\ref{warunek_na_typ_D})). After some algebraic work we find that the condition (\ref{warunek_na_typ_D}) under (\ref{ASD_curvature_nonexpanding}) reduces to two equations
\begin{equation}
\label{rownania_na_watunek_typu_D}
\frac{\partial}{\partial y}  \left( \frac{\partial}{\partial p} \frac{\delta_{pp}}{\delta_{p}} - \frac{1}{2} \frac{\delta_{pp}^{2}}{\delta_{p}^{2}} \right) = 0 \ , \ \ \ \frac{\partial}{\partial p}  \left( \frac{\partial}{\partial y} \frac{\beta_{yy}}{\beta_{y}} - \frac{1}{2} \frac{\beta_{yy}^{2}}{\beta_{y}^{2}} \right) = 0
\end{equation}
With the help of gauge freedom (\ref{gauge_freedom_po_uproszczeniach}), Eqs. (\ref{rownania_na_watunek_typu_D}) can be easily solved and the solutions read
\begin{equation}
\label{funkcje_betaidelta_dlatypu_DxD}
\beta(y,p) = \frac{\beta_{1}(p)}{y+ \beta_{2}(p)} \ , \ \ \ \delta(y,p) = \frac{\delta_{1}(y)}{p+ \delta_{2}(y)}
\end{equation}
where $\beta_{i}$ and $\delta_{i}$ ($i=1,2$) are arbitrary functions of their variables. 

[\textbf{Remark.} The metric of the type $[\textrm{D}]^{nn} \otimes [\textrm{D}]^{nn}$ given by (\ref{para_Kahler_nonexpandingASD_nullstrings}) with (\ref{funkcje_betaidelta_dlatypu_DxD}) have been analyzed recently by Paweł Nurowski. Investigating the symmetries of this solution Nurowski proved that the metric is equivalent to the following one \cite{Nurowski}
\begin{equation}
\label{Klein_Beltrami_metric}
ds^{2}_{N} = \frac{2dpdx}{(1+\frac{\Lambda}{2} px)^{2}} + \frac{2dqdy}{(1+\frac{\Lambda}{2}qy)^{2}}
\end{equation}
The metric (\ref{Klein_Beltrami_metric}) is an interesting example of para-K\"{a}hler Einstein structure: it is equivalent to locally symmetric model $\mathcal{M}=\textbf{SO}(2,2) / \textrm{maximal torus}$. It corresponds to the homogeneous para-K\"{a}hler structure.

In conclusion, we see that the functions $\beta_{i}$ and $\delta_{i}$ can be brought to the form $\beta_{1}=-1$, $\beta_{2}=0$, $\delta_{1}=-1$ and $\delta_{2}=0$ (although it is hard to find the explicit transformation, which brings those functions to such forms). Then $\beta=-y^{-1}$ and $\delta=-p^{-1}$. Changing the coordinates $\Lambda p':=2p$ and $\Lambda q':=2q$ and dropping primes we find, that the metric (\ref{para_hermite_dwiewstegi_nonexpanding}) generated by (\ref{para_Kahler_nonexpandingASD_nullstrings}) with $\beta=-y^{-1}$ and $\delta=-p^{-1}$ gives exactly the metric (\ref{Klein_Beltrami_metric}).]

\subsubsection{Expanding case}

Consider now the case with expanding congruence of ASD null strings. It means that $r_{y} \ne 0$. Introducing new functions $z=z(y,p,q)$ and $G=G(y,p,q)$ such that
\begin{equation}
r=\frac{z_{q}}{z_{p}} \ , \ \ \ s = G_{q} - \frac{z_{q}}{z_{p}} \, G_{p}
\end{equation}
one finds the general solution of Eq. (\ref{warunki_przy_istnieniu_ASD_struny}) in the form
\begin{equation}
M(p,q,x,y) = F(x,y,z(y,p,q)) +  G(y,p,q)
\end{equation}
Inserting this into the Einstein equations (\ref{Einstein_equations_for_Dxany_non}) we find that 
\begin{equation}
G = -\frac{1}{\Lambda} \ln (z_{p}z_{qy}-z_{q}z_{py})
\end{equation}
The Einstein equations reduce to
\begin{equation}
\frac{1}{F_{zx}} e^{-\Lambda F} = F_{z} + \beta \ , \ \ \ \beta (y,z) := \frac{z_{p}G_{qy}-z_{q}G_{py}}{z_{p}z_{qy}-z_{q}z_{py}}
\end{equation}
and the equation (\ref{consequences_of_Cdot4zero}) is identically satisfied. 

Unfortunately, we have not been able to find any solution of this system, even for the types $[\textrm{D}]^{nn} \otimes [\textrm{III,N}]^{e}$ (i.e., under the additional condition $\dot{C}^{(3)}=0$ or $\dot{C}^{(3)}=\dot{C}^{(2)}=0$) or under \textsl{ad hoc} assumption $\beta=0$. It proves, somehow surprisingly, that the coordinate system $(x,y,p,q)$ introduced in (\ref{para_Kahler_null_tetrad}) is not the most suitable for our purpose. Nevertheless, we can attack the problem using machinery of nonexpanding hyperheavenly spaces theory.

[\textbf{Remark}. We should mention, however, the simple solution which belongs to the class with expanding  congruence of ASD null strings. It is generated by the function
\begin{equation}
M = \frac{3}{\Lambda} \ln \left( p +qx + \frac{\Lambda^{2}}{9}y \right)
\end{equation}
The only nonzero curvature coefficient is $C^{(3)} = -\frac{2\Lambda}{3}$ so this solution is of the type $[\textrm{D}]^{nn} \otimes [-]^{e}$. Straightforward calculations give the metric, which after changing the names of the coordinates, $p \rightarrow b$, $q \rightarrow a$, $y \rightarrow -9y / \Lambda^{2}$ assumes the form
\begin{equation}
\label{dancing_metric}
ds^{2} = - \frac{6}{\Lambda (b+ax-y)^{2}} \Big( ((y-b)dx-xdy)da + (adx-dy)db \Big)
\end{equation}
The metric (\ref{dancing_metric}) is exactly the \textsl{dancing metric} \cite{Nurowski_Bor_Lamoneda}. We are indebted to Paweł Nurowski who expressed this solution in double null coordinates formalism \cite{Nurowski}.]

%#####################################################################################

\section{Para-K\"{a}hler spaces in Plebański - Robinson - Finley coordinates}
\label{para_Kahler_PRF}
\setcounter{equation}{0}

\subsection{The metric of the nonexpanding hyperheavenly spaces of the types $[\textrm{D}]^{nn} \otimes [\textrm{any}]$}
\label{para_Kahler_PRF_general_description}

The hyperheavenly space theory has been developed in many papers (e.g. see \cite{Plebanski_Finley, Chudecki_1}). We only mention here that in coordinate system $(x,y,p,q)$ the metric of $\mathcal{HH}$ spaces of the types $[\textrm{D}]^{nn}  \otimes [\textrm{any}]$ endowed with two complementary, nonexpanding congruences of SD null strings can be brought to the form
\begin{eqnarray}
\label{nonexpanding_metric_HHspaces}
ds^{2} &=& 2dydq - 2dxdp + 2 \Big( \frac{\Lambda}{3} x^{2}- \Theta_{yy} \Big) dp^{2} + 2 \Big( \frac{\Lambda}{3}y^{2}-\Theta_{xx} \Big) dq^2 
\\ \nonumber
&& - 4 \Big( \frac{\Lambda}{3}xy+\Theta_{xy} \Big) dpdq
\end{eqnarray}
The null tetrad which seems to be the most suitable for our purposes (Plebański-Robinson-Finley tetrad) reads
\begin{eqnarray}
&& e^{1}=-dq \ , \ \ \ e^{2} = -dy + \Big( \frac{\Lambda}{3}xy+\Theta_{xy} \Big) dp - \Big( \frac{\Lambda}{3} y^{2} - \Theta_{xx} \Big) dq
\\ \nonumber
&& e^{3}=dp \ , \ \ \ e^{4} = -dx + \Big( \frac{\Lambda}{3}x^2-\Theta_{yy} \Big) dp - \Big( \frac{\Lambda}{3} xy + \Theta_{xy} \Big) dq
\end{eqnarray}
One finds the connection forms to be
\begin{eqnarray}
\label{formy_konneksji_dla_HHnonexp}
&&\Gamma_{42}=\Gamma_{31}=0 \ , \ \ \ \Gamma_{12}+\Gamma_{34}=\Lambda (ydq-xdp)
\\ \nonumber
&& \Gamma_{41} = \Theta_{xxx}dq + \Theta_{xxy}dp + \frac{\Lambda}{3}ydp
\\ \nonumber
&& \Gamma_{32} = \Theta_{yyx}dq + \Theta_{yyy}dp + \frac{\Lambda}{3}xdq
\\ \nonumber
&&-\Gamma_{12} + \Gamma_{34} = 2\Theta_{xxy}dq + 2 \Theta_{xyy} dp- \frac{\Lambda}{3}ydq - \frac{\Lambda}{3}xdp
\end{eqnarray}
and the curvature is given by
\begin{eqnarray}
&& C^{(5)}=C^{(4)}=C^{(2)}=C^{(1)}=0 \ , \ \ \ C^{(3)}=-\frac{2}{3}\Lambda
\\ \nonumber
&& \dot{C}^{(5)} = 2 \Theta_{xxxx} \ , \  \dot{C}^{(4)} = 2 \Theta_{xxxy} \ ,  \ \dot{C}^{(3)} = 2 \Theta_{xxyy} \ , \  \dot{C}^{(2)} = 2 \Theta_{xyyy} \ ,  \ \dot{C}^{(1)} = 2 \Theta_{yyyy} 
\end{eqnarray}
The vacuum Einstein equations reduce to the \textsl{nonexpanding hyperheavenly equation with $\Lambda$}
\begin{equation}
\Theta_{xx}\Theta_{yy} - \Theta_{xy}^{2} + \Theta_{yq}-\Theta_{xp} + \frac{\Lambda}{3} ( 3x\Theta_{x} + 3y\Theta_{y}-3\Theta - x^{2} \Theta_{xx} -y^{2} \Theta_{yy} -2xy \Theta_{xy} )=0
\end{equation}
The condition for algebraic degeneration of the ASD side is once again very involved. This is why we assume that ASD null string lies in the plane $(\partial_{1}, \partial_{4})$ what is equivalent to the conditions $\Gamma_{411}=\Gamma_{414}=0$. From (\ref{formy_konneksji_dla_HHnonexp}) we find $\Theta_{xxx}=0$ and, consequently, the key function is
\begin{equation}
\Theta = \frac{1}{2} A(y,p,q) x^{2} + B(y,p,q) x + C(y,p,q)
\end{equation}
Hyperheavenly equation splits into the set of three equations
\begin{subequations}
\begin{eqnarray}
\label{non_hyp_1}
&& 3AA_{yy} - 6A_{y}^{2}+ 3A_{yq}+ \Lambda (A-yA_{y}-y^{2}A_{yy}) = 0
\\ 
\label{non_hyp_2}
&& 3AB_{yy} - 6A_{y}B_{y} + 3B_{yq} - 3A_{p} + \Lambda (yB_{y}-y^{2}B_{yy}) = 0
\\
\label{non_hyp_3}
&& AC_{yy} - B_{y}^{2}+C_{yq}-B_{p} + \frac{\Lambda}{3} (3yC_{y} - 3C-y^{2}C_{yy})=0
\end{eqnarray}
\end{subequations}
The expansion of the ASD congruence is given by $\Gamma_{423}=A_{y}+ \frac{\Lambda}{3}y$.

\subsection{Hyperheavenly spaces of the types $[\textrm{D}]^{nn} \otimes [\textrm{II}]^{n}$ and $[\textrm{D}]^{nn} \otimes [\textrm{D}]^{nn}$}
\label{subsekcja_hyperheavenly_DxIID}

In this subsection we investigate the spaces equipped with two nonexpanding congruences of SD null strings and one (types $[\textrm{D}]^{nn} \otimes [\textrm{II}]^{n}$) or two (types $[\textrm{D}]^{nn} \otimes [\textrm{D}]^{nn}$)  nonexpanding congruences of ASD null strings. Taking $\Gamma_{423}=0$ (ASD null string is nonexpanding), we find from (\ref{non_hyp_1}) the solution for $A$
\begin{equation}
\label{solution_for_A}
A=-\frac{\Lambda}{6} y^{2}
\end{equation}
(The hyperheavenly gauge freedom has been used to bring $A$ to the form (\ref{solution_for_A}). In what follows this gauge is extensively employed. This involves somehow tedious calculations which are not presented here. For detailed discussion about (nonexpanding) hyperheavenly spaces gauge freedom, see \cite{Plebanski_Finley, Chudecki_3}). Then the solution of (\ref{non_hyp_2}) reads
\begin{equation}
\label{solution_for_B}
B_{y} = y^{2} \, f(p,z) \ , \ \ \ z:= q-\frac{2}{\Lambda y}
\end{equation}
Differentiating Eq. (\ref{non_hyp_3}) once by $\partial_{y}$ we arrive at the equation from which we find the solution for $C_{yy}$
\begin{equation}
C_{yy} = h(p,z) - \frac{4}{\Lambda} y^{2} f^{2} - \frac{8}{\Lambda^{2}}y ff_{z} - \frac{2}{\Lambda}y f_{p}
\end{equation}
Functions $f=f(p,z)$ and $h=h(p,z)$ are arbitrary functions of their variables. [Note that because of arbitrariness of function $f$, it is impossible to find the explicit form of $B$ from equation (\ref{solution_for_B}). Similarly, $C_{yy}$ cannot be explicitly integrated, so the complete solution of the system  (\ref{non_hyp_1})-(\ref{non_hyp_3}) remains unknown. However, the hyperheavenly metric is determined with the precision up to the second derivatives of the key function $\Theta$, so $A$, $B_{y}$ and $C_{yy}$ contain the entire information needed to obtain the explicit form of the metric]. 

Finally, passing from coordinates $(p,q,x,y)$ to $(p,z,x,y)$ we find the metric in the form
\begin{eqnarray}
\label{Metryka_DxII_nieeks}
ds^{2} &=& -2dxdp -2dydz+\Lambda y^{2}dz^{2} -4y^{2}f \Big( dz - \frac{2}{\Lambda y^{2}} dy \Big) dp
\\ \nonumber
&& +2 \bigg( \frac{\Lambda x^{2}}{2} -2x \Big(yf+ \frac{f_{z}}{\Lambda } \Big) -h + \frac{4}{\Lambda}f^{2}y^{2} + \frac{8}{\Lambda^{2}}ff_{z}y + \frac{2}{\Lambda}yf_{p} \bigg) dp^{2}
\end{eqnarray}
and the curvature coefficients are
\begin{eqnarray}
&&C^{(3)}=\dot{C}^{(3)}=-\frac{2}{3}\Lambda \ , \ \ \ \dot{C}^{(2)} = 4f +\frac{8f_{z}}{\Lambda y} + \frac{8f_{zz}}{\Lambda^{2}y^{2}}
\\ \nonumber
&& \dot{C}^{(1)} = \frac{16}{\Lambda^{3}y^{4}} f_{zzz} \, x + \frac{8}{\Lambda^{2} y^{4}} (h_{zz}-\Lambda y h_{z}) - \frac{16}{\Lambda}f^{2}- \frac{64}{\Lambda^{2}y} ff_{z}
\\ \nonumber
&& \ \ \ \ \ \ \ 
- \frac{16}{\Lambda^{3} y^{3}} \bigg(   4y(f_{z}^{2}+ff_{zz}) + \frac{4}{\Lambda} (3f_{z}f_{zz}+ff_{zzz}) +  f_{pzz}  \bigg)
\end{eqnarray}
The condition for the ASD type $[\textrm{D}]$ (\ref{warunek_na_typ_D}) leads to two equations for the functions $f$ and $h$. However, using coordinate gauge freedom one proves that both these functions can be brought to zero without any loss of generality. With $f=h=0$ the metric of the type $[\textrm{D}]^{nn} \otimes [\textrm{D}]^{nn}$ has the form
\begin{equation}
\label{Metryka_DxD_nieeks}
ds^{2} = -2dxdp -2dydz + \Lambda y^{2}dz^{2} + \Lambda x^{2}dp^{2} 
\end{equation}
Then changing the variables $(x,y,p,q) \rightarrow (x',y',p',q')$ according to
\begin{equation}
x=-\frac{2}{\Lambda} \frac{x'}{1+\frac{\Lambda}{2}p'x' } \ , \ \ \ y=-\frac{2}{\Lambda} \frac{y'}{1+\frac{\Lambda}{2}z'y'} \ , \ \ \ p=\frac{\Lambda}{2} p' \ , \ \ \ z=\frac{\Lambda}{2}z'
\end{equation}
and dropping primes we get the metric (\ref{Klein_Beltrami_metric}).

\subsection{Hyperheavenly spaces of the types $[\textrm{D}]^{nn} \otimes [\textrm{III,N}]^{e}$}

Now we consider the spaces of the types $[\textrm{D}]^{nn} \otimes [\textrm{III,N}]^{e}$ with $\Lambda \ne 0$. In this case ASD null string is necessarily expanding (compare Theorem \ref{twierdzenie_o_wstegach_nieekspandujacych}). The condition $\dot{C}^{(3)}=0$ gives $A_{yy}=0$, but the function $A$ can be gauged to zero. Then we proceed analogously as  in subsection \ref{subsekcja_hyperheavenly_DxIID}. Finally, we arrive at the following results
\begin{eqnarray}
&& A=0 \ , \ \ \ B_{y}=y \, f(p,z) \ , \ \ \ z:=q-\frac{3}{\Lambda y}
\\ \nonumber
&& C_{yy}=y \, h(p,z) + \frac{9ff_{z}}{\Lambda^{2}y} + \frac{6f^{2}}{\Lambda} + \frac{3f_{p}}{\Lambda}
\end{eqnarray} 
where $f$ and $h$ are arbitrary functions of their variables. Taking $(p,q,x,y) \ \rightarrow \ (p,z,x,y)$ we find the metric in the form
\begin{eqnarray}
\label{metryka_typow_DxIIIN}
ds^{2} &=& -2dxdp+ \frac{2\Lambda}{3}y^{2} \Big( dz - \frac{3}{\Lambda y^{2}} dy \Big) dz -4 \Big(yf+\frac{\Lambda}{3}xy \Big) \Big( dz - \frac{3}{\Lambda y^{2}} dy \Big) dp \ \ \ \ \ \ 
\\ \nonumber
&& -2 \bigg(  yh+x \Big(f+ \frac{3f_{z}}{\Lambda y} \Big) - \frac{\Lambda}{3} x^{2}  + \frac{9ff_{z}}{\Lambda^{2}y} + \frac{6}{\Lambda} f^{2} + \frac{3}{\Lambda} f_{p} \bigg) dp^{2}
\end{eqnarray}
and the curvature reads
\begin{eqnarray}
&&C^{(3)}=-\frac{2}{3}\Lambda \ , \ \ \ \dot{C}^{(2)} = \frac{18 f_{zz}}{\Lambda^{2}y^{3}}
\\ \nonumber
&& \dot{C}^{(1)} = \frac{54}{\Lambda^{3}y^{5}} (f_{zzz}-\Lambda y f_{zz}) \, x + \frac{18h_{zz}}{\Lambda^{2}y^{3}} 
\\ \nonumber
&& \ \ \ \ \ \ \ \ \ 
+ \frac{18}{\Lambda^{2}} \bigg( \frac{-6ff_{z}}{y^{3}} + \frac{27f_{z}f_{zz}+9ff_{zzz}}{\Lambda^{2}y^{5}} - \frac{2f_{pz}}{y^{3}} + \frac{3f_{pzz}}{\Lambda y^{4}} \bigg)
\end{eqnarray}
Type $[\textrm{D}]^{nn} \otimes [\textrm{N}]^{e}$ can be obtained from the metric (\ref{metryka_typow_DxIIIN}) by demanding that $\dot{C}^{(2)}=0$. It is equivalent to the condition $f_{zz}=0$ but the gauge freedom still available allows one to bring the function $f$ into the form $f=f_{0}z$ where $f_{0}$ is an arbitrary constant.

\subsection{Hyperheavenly spaces of the types $[\textrm{D}]^{nn} \otimes [-]^{e}$}

Finally, we mention that all solutions of the hyperheavenly spaces of the type  $[\textrm{D}]^{nn} \otimes [-]^{e}$ with $\Lambda \ne 0$ are known. They belong to the SD Einstein Walker spaces and are presented, for example, in \cite{Chudecki_1} (the metric (4.2) with (4.29) in Ref. \cite{Chudecki_1}). This solution depends on 6 arbitrary functions of two variables and it is clear that some of these functions can be gauged away. It is an interesting question what is the simplest form of the metric of the type $[\textrm{D}]^{nn} \otimes [-]^{e}$ with $\Lambda \ne 0$ but we do not enter this question here.

%#####################################################################################

\section{Concluding Remarks}

In the present paper we have dealt with para-Hermite and para-K\"{a}hler Einstein spaces with $\Lambda \ne 0$. The existence of two complementary congruences of SD null strings  has allowed us to choose the coordinates in such a manner that we have been able to obtain the explicit examples of the spaces of the types $[\textrm{D}]^{ee} \otimes [\textrm{deg}]$ and $[\textrm{D}]^{nn} \otimes [\textrm{deg}]$. Such spaces are interesting from the geometrical point of view. Since we also assume the existence of a congruence of ASD null strings all the metrics presented in the paper are two-sided Walker and two-sided sesqui Walker spaces \cite{Chudecki_1,Law_1}. Moreover, para-K\"{a}hler Einstein spaces have just found their place in geometrical approach to the mechanical issues \cite{Nurowski_An,Nurowski_Bor_Lamoneda,Nurowski}. 

In the Table \ref{Podsumowanie_metryk} we listed all possible types of para-Hermite and para-K\"{a}hler Einstein spaces with $\Lambda \ne 0$ together with the metrics which have been found in our paper. It is clear that yet no examples of four classes of the metrics have been found. We believe however, that there exists the coordinate frame more suitable for our purpose which enables us to find the solutions.

\begin{table}[h]
\begin{center}
\begin{tabular}{|c|c|c|}   \hline
  & $[\textrm{D}]^{ee}$ (SD para-Hermite) & $[\textrm{D}]^{nn}$ (SD para-K\"{a}hler) \\ \hline
$[\textrm{I}]$       & not found & not found \\ \hline
$[\textrm{II}]^{e}$  & (\ref{metrykka_typu_DxII_1}) with $f\ne0$ &  not found \\ 
                     & (\ref{metrykka_typu_DxII_2}) with $f_{y}\ne0$ & \\ \hline
$[\textrm{II}]^{n}$  & (\ref{metrykka_typu_DxII_1}) with $f=0$  & (\ref{Metryka_DxII_nieeks}) \\ \hline
$[\textrm{D}]^{ee}$  & (\ref{Metryka_DxD_clasa1}) with $S_{0} \ne 0$ & after changing orientation the same \\ 
                     & (\ref{metryka_typu_DD_class2_a}) with $S_{0} \ne 0$ & metrics as $[\textrm{D}]^{ee} \otimes [\textrm{D}]^{nn}$ \\ \hline
$[\textrm{D}]^{nn}$  & (\ref{Metryka_DxD_clasa1}) with $S_{0} = 0$ & (\ref{Metryka_DxD_nieeks}) \\ 
                     & (\ref{metryka_typu_DD_class2_a}) with $S_{0} = 0$ &  \\ \hline
$[\textrm{III}]^{e}$ & (\ref{metryka_DxIII_class1}) & (\ref{metryka_typow_DxIIIN}) \\ \hline
$[\textrm{III}]^{n}$ & do not exist & do not exist \\ \hline
$[\textrm{N}]^{e}$   & not found & (\ref{metryka_typow_DxIIIN}) with $f=f_{0}z$ \\ \hline
$[\textrm{N}]^{n}$   & do not exist & do not exist \\ \hline
\end{tabular}
\caption{Metrics of para-Hermite and para-K\"{a}hler Einstein spaces with $\Lambda \ne 0$.}
\label{Podsumowanie_metryk}
\end{center}
\end{table}

It is worth to pointing out once again that all our solutions are only of special classes. Some (unpublished yet) results found by Nurowski contain new, but still special metrics of the types $[\textrm{D}]^{nn} \otimes [\textrm{II,N}]$ \cite{Nurowski}. Those metrics provide us with explicit examples of para-K\"{a}hler Einstein spaces with $\Lambda \ne 0$ other then those presented in our paper. 
\newline
\newline
\textbf{Acknowledgments}
\newline
Some points of this paper were presented in November 2015 at the \textsl{Second Conference of Polish Society on Relativity: 100 Years of General Relativity}. The author is indebted to Maciej Przanowski and Paweł Nurowski for many illuminating discussions and is especially grateful to Michał Dobrski for his interest in this work and the help in a few crucial matters. Some useful suggestions of the anonymous referee are highly appreciated.

%#####################################################################################

\end{document}